\documentclass[10pt,journal,compsoc]{IEEEtran}
\usepackage{mypreamble}
\usepackage{relsize}
\usepackage[inline]{enumitem}
\usepackage{tablefootnote}
\usepackage{makecell}

\newcommand{\revise}[2]{{\color{black}{#2}}}
\newcommand\fix[1]{{\textcolor{black}{#1}}}

\newcommand\our{PyCoder}
\newcommand\ea[1]{\emph{et al.~}}
\usepackage{tcolorbox}
\newcommand*\circled[1]{\tikz[baseline=(char.base)]{
            \node[shape=circle,draw,inner sep=0.5pt] (char) {\small{#1}};}}
\usepackage{amsmath, nccmath}

\DeclareMathOperator*{\argmin}{argmin}

\newcommand\rqone{What is the performance of our \our~for the token-level and line-level code completion tasks when compared to state-of-the-art models?}
\newcommand\rqtwo{What is the impact of the training strategies on the performance of our \our?}
\newcommand\rqthree{What is the impact of the task weighting parameters in multi-task learning on the performance of our \our?}
\newcommand\rqfour{What is the impact of the decoding methods on the performance of our \our?}

\newcommand\fdone{
\our~achieves the first rank on the CodeXGLUE leaderboard with an accuracy of 77.12\% for the token-level predictions, which is 0.43\%-24.25\% more accurate than  baselines.
In addition, \our~achieves an exact match of 43.37\% for the line-level predictions, which is 3.63\%-84.73\% more accurate than baselines.
}
\newcommand\fdtwo{
Multi-task training strategies have an impact on \our~for both token-level and line-level predictions.
We find that \our-Hard performs best; followed by \our-IFN and \our-Soft.
}
\newcommand\fdthree{\our~is generally robust to the task weighting parameters, achieving comparative (without task weighting) or better (with task weighting) performance when compared to the baselines.}
\newcommand\fdfour{Decoding methods have an impact on the performance of \our~with an exact match varying from 33.80\% to 41.52\% for line-level predictions. 
Beam Search performs best, while Sampling performs worst.
}

\begin{document}

%! Author = sbbfti
%! Date = 10/06/2020
% \newacronym{ts}{$T_{s}$}{supply air temperature, $^{\circ}$C}
% \newacronym{vsp}{$\dot{V}_{sp}$}{volumetric flow rate set-point, L/s}

% \newacronym{gtt}{GTT}{Generative Type-Aware Transformers}
\newacronym{mtl}{MTL}{Multi-Tasks Learning}
\newacronym{ifn}{STILTs}{Intermediate Fine-tuning}
% {Supplementary Training on Intermediate Labeled-data Tasks Model}
\newacronym{em}{EM}{Exact Match}
\newacronym{es}{ES}{Edit Similarity}
\newacronym{mrr}{MRR}{Mean Reciprocal Rank}
\newacronym{bpe}{BPE}{Byte Pair Encoding \cite{sennrich2015neural}}
\newacronym{nlp}{NLP}{natural language processing}
\newacronym{ide}{IDEs}{Integrated Development Environments}
\newacronym{ast}{ASTs}{Abstract Syntax Trees}

\title{Syntax-Aware On-the-Fly Code Completion}
% \title{Syntax-Aware On-the-Fly Python \\Code Completion with Multi-Task Learning}

\author{Wannita~Takerngsaksiri,~\IEEEmembership{Student Member,~IEEE,}
        Chakkrit Tantithamthavorn,~\IEEEmembership{Member,~IEEE,}
        and~Yuan-Fang Li,~\IEEEmembership{Member,~IEEE.}% <-this % stops a space
\IEEEcompsocitemizethanks{\IEEEcompsocthanksitem W. Takerngsaksiri, C. Tantithamthavorn, Y.-F. Li are with  the Faculty of Information Technology, Monash University, Australia.\protect\\
E-mail: \{wannita.takerngsaksiri, chakkrit, yuanfang.li\}@monash.edu
% \IEEEcompsocthanksitem J. Doe and J. Doe are with Anonymous University.
}% <-this % stops a space
\thanks{Manuscript received November 4, 2022, revised April 6, 2023.}}
% 

\markboth{IEEE Transactions on Software Engineering (TSE)}%
{Takerngsaksiri \MakeLowercase{\textit{et al.}}: Syntax-Aware On-the-Fly Code Completion}
% \markboth{Journal of \LaTeX\ Class Files,~Vol.~14, No.~8, August~2015}%
% {Shell \MakeLowercase{\textit{et al.}}: Bare Advanced Demo of IEEEtran.cls for IEEE Computer Society Journals}

\IEEEtitleabstractindextext{%
\begin{abstract}

Code completion aims to help improve developers' productivity by suggesting the next code tokens from a given context. 
Various approaches have been proposed to incorporate abstract syntax tree (AST) information for model training, ensuring that code completion is aware of the syntax of the programming languages.
However, existing syntax-aware code completion approaches are not on-the-fly, as we found that for every two-thirds of characters that developers type, AST fails to be extracted because it requires the syntactically correct source code, limiting its practicality in real-world scenarios.
On the other hand, existing on-the-fly code completion does not consider syntactic information yet.
In this paper, we propose \our~to leverage token types, a kind of lightweight syntactic information, which is readily available and aligns with the natural order of source code.
Our \our~is trained in a multi-task training manner so that by learning the supporting task of predicting token types during the training phase, the models achieve better performance on predicting tokens and lines of code without the need for token types in the inference phase.
Comprehensive experiments show that \our~achieves the first rank on the CodeXGLUE leaderboard with an accuracy of 77.12\% for the token-level predictions, which is 0.43\%-24.25\% more accurate than baselines.
In addition, \our~achieves an exact match of 43.37\% for the line-level predictions, which is 3.63\%-84.73\% more accurate than baselines.
These results lead us to conclude that token type information (an alternative to syntactic information) that is rarely used in the past can greatly improve the performance of code completion approaches, without requiring the syntactically correct source code like AST-based approaches do.
Our \our~is publicly available on HuggingFace and GitHub.

\end{abstract}

}

\maketitle

% \linenumbers

% Document

\IEEEdisplaynontitleabstractindextext
\IEEEpeerreviewmaketitle
\ifCLASSOPTIONcompsoc
\IEEEraisesectionheading{\section{Introduction}\label{sec:introduction}}
\else
\section{Introduction}
\label{sec:introduction}
\fi

\IEEEPARstart{C}{ode} completion, or AutoCompletion, is one of the most essential features in modern \gls{ide} (e.g., GitHub's Copilot, Intellisense in Visual Studio Code \cite{intelliSense}).
% Tabnine\footnote{https://www.tabnine.com}
The goal of code completion is to automatically recommend source code based on a given context, which could help developers reduce the amount of typing and coding iteration time and eliminate the number of typo errors.
A recent study conducted by Google found that the current code completion feature could reduce developers' effort by 6\% and context switching by 7\%~\cite{ml2022google}.

% the next source code statements given the current context. It can help improve developers productivity by minimize the amount of typing and eliminate typos. These may include pop-up when typing, suggest methods or objects, or query parameters of functions.
% \kla{better to mention the existing tools too, e.g., intellisense, tabnine}\
% \kla{also better to mention the benefit https://ai.googleblog.com/2022/07/ml-enhanced-code-completion-improves.html / check this link and put the reference.}
% Some existing tools are, for example, the code completion in IntelliSense \cite{intelliSense} which is a part of code editing features in Visual Studio Code (VS Code), and Tabnine \footnote{https://www.tabnine.com} which is the AI code assistant for whole-line and full-function code completion that can integrate to many \gls{ide} platforms.

% \kla{2nd para should start from recent work, not the history. this para can be in the related work.}

Recent code completion approaches often leverage modern deep learning architectures (e.g., Recurrent Neural Network, Transformer architecture) to exploit their strong representation power.
More specifically, state-of-the-art code completion models (e.g., CodeGPT~\cite{lu2021codexglue}, GPT-2~\cite{radford2019language}, GPT-C~\cite{svyatkovskiy2020intellicode}, TavTrans~\cite{kim2021code}, CodeFill~\cite{izadi2022codefill}) are based on code-focused large language models (LLMs) that are trained from large codebase and natural language corpora (e.g., the CodeSearchNet corpus with 2 million GitHub repositories). These LLMs are fine-tuned on a specific dataset to perform specific tasks (e.g., code completion).
However, existing code completion approaches have the following limitations.

\textbf{Limitation 1: \emph{On-the-fly code completions approaches do not consider syntactic information.}}
On-the-fly code completion approaches are designed to generate code tokens based on a given context without requiring the completeness of prior context.
Represent techniques include GPT-2~\cite{radford2019language}, a Transformer-based decoder model for generative tasks pre-trained on English webpage datasets; CodeGPT~\cite{lu2021codexglue}, a GPT-2 model architecture pre-trained on source code datasets; and GPT-C~\cite{svyatkovskiy2020intellicode}, a GPT-2 model architecture pre-trained on multi-language source code. In their pre-training, these models learn to complete the next code tokens. 
In doing so, the performance of these on-the-fly code completion approaches is limited by their lack of consideration of syntactic information.
% provided in the source code.

\textbf{Limitation 2: \emph{Existing syntax-aware code completion approaches are not on-the-fly.}}
To ensure that the generated source code is syntactically correct~\cite{brockschmidt2018generative}, researchers proposed to leverage the Abstract Syntax Tree (AST) information~\cite{li2017code, svyatkovskiy2019pythia, kim2021code, liu2020self, izadi2022codefill, liu2022unified}.
For example, Kim~\ea~\cite{kim2021code} proposed TravTrans, 
% \kla{not clear yet, tell what is this approach}
a Transformer-based architecture consuming syntactic information from a variety representations of ASTs traversal;
Izadi~\ea~\cite{izadi2022codefill} proposed CodeFill, a multi-task,  
% \kla{parallel is a new jargon that is never clearly defined. avoid using it.}
Transformer-based architecture consuming source code and AST types. 
While existing AST-based code completion approaches may generate code that is more syntactically correct, the application scenario remains limited.
In particular, the existing AST-based code completion approaches~\cite{li2017code, svyatkovskiy2019pythia, kim2021code, liu2020self, izadi2022codefill, liu2022unified} require source code to be completed (i.e., all the previous tokens are valid and parsable) at the inference time so the AST information can be obtained from the source code.
However, our motivating analysis found that in practice, two thirds of the source code characters is incomplete and not parsable (e.g., containing syntax errors), making the existing AST-based code completion approaches \emph{inapplicable} in real-world scenarios.

\emph{In this paper}, we propose \our, an automated code completion approach that can generate source code at any time regardless of the completeness of the source code, i.e., \textbf{syntax-aware on-the-fly code completion}.
Our approach is designed to consider the syntactic information of the source code during the learning phase, but \emph{does not} require syntactic information during the inference phase.
Instead of using the AST information like in previous works~\cite{kim2021code, izadi2022codefill, li2017code, svyatkovskiy2019pythia, liu2020self, liu2022unified}, we propose to leverage the \emph{token type} information (e.g., String, Number, Name, Keyword), which is a readily-available and light-weight syntactic information without requiring the completeness of the source code.
During the learning process, we design our approach to carry out two prediction tasks, i.e., the token prediction task and the type prediction task.
% Unlike some previous work that leverages AST information as an individual learning objective for the AST node prediction task~\cite{kim2021code,li2017code},
% Our research study on how to best use this token type information along with source code.
To ensure that our model captures both syntactic and semantic information during the training process, we leverage Multi-Task Training (MTT) techniques to learn both the token prediction task and the type prediction task. 
% prediction tasks, and intermediate fine-tuning (STILTs) that sequentially learns from task-to-task.
% To address this challenge of training on multiple tasks (i.e., code prediction and type prediction tasks),
% Particularly, 
Given a sequence of code tokens, our approach performs the following steps: (1) extract the token type information of each token, (2) perform the sub-word tokenization on each token, (3) align token type data with sub-word source code data, and (4) build a code completion model using a GPT-2 architecture based on a pre-trained CodeGPT language model with a multi-task training technique.
% \kla{highlight we don't use the type information at the inference time.}

In our experiment, we compare our \our~with \revise{R3.15}{five} existing state-of-the-art models (i.e., Pointer Mixture Network~\cite{li2017code}, TravTrans~\cite{kim2021code}, GPT-2~\cite{radford2019language}, CodeGPT~\cite{lu2021codexglue}, and \fix{UniXcoder~\cite{guo2022unixcoder}}).
During the inference phase, we evaluate our approach based on the token-level and line-level prediction tasks.
Through an extensive evaluation on the \textit{PY150}~\cite{raychev2016probabilistic} standard benchmark Python dataset for the code completion task that is used in Microsoft's CodeXGLUE benchmark~\cite{lu2021codexglue}, we answer the following research questions:

% \kla{revised till here}

\begin{description}[leftmargin=1cm, itemsep=3pt]
  \item[RQ1)] \textbf{\rqone}\\
  \textbf{Results.} \fdone

% \our~achieves the first rank on the CodeXGLUE leaderboard with an accuracy of 77.12\% for the token-level predictions, which is 0.43\%-24.25\% more accurate than other baselines.
% In addition, \our~achieves an exact match of 43.37\% for the line-level predictions, which is 3.63\%-84.73\% more accurate than other baselines.

%   Compared with second-best models, \our~achieves an improvement on accuracy of 1.89\% (from 75.69\% to 77.12\%) in token-level prediction and on exact match of 8.34\% %3.34\% 
%   \kla{use relative percentage} 
%   (from 40.03\% to 43.37\%) in line-level prediction.
  
  \item[RQ2)] \textbf{\rqtwo}\\
  \textbf{Results.} \fdtwo

%   the different training techniques can vary accuracy by 2.98\% %2.23\% (from 74.77\% to 77.00\%) in token-level prediction, and 
%   the exact match is different by 11.86\% (i.e., max-min) %4.54\% (from 38.29\% to 42.83\%) in line-level prediction.
  
  \item[RQ3)] \textbf{\rqthree}\\
  \textbf{Results.} \fdthree
%   When adjusting the weights (section~\ref{sec:approach-weight}) of the two tasks (token type prediction and token value prediction), accuracy in token-level prediction can differ by 2.21\% %1.67\% 
%   (from 75.45\% to 77.1\%) and exact match in line-level prediction can differ by 9.05\% %3.6\%
%   (from 39.77\% to 43.37\%).
%   The results show that \our~can perform better in both token-level and line-level predictions when tuning the task weighing parameters.
  
  \item[RQ4)] \textbf{\rqfour}\\
  \textbf{Results.} \fdfour
%   The different decoding methods can vary the results in line-level prediction by 22.84\% %7.72\% 
%   (from 33.80\% to 41.52\%) on exact match and 7.65\% %5.26\%
%   (from 68.72\% to 73.98\%) on edit similarity.
%   Thus, decoding methods have a significant impact on the quality of predictions, and sampling-based methods may not be suitable for the code completion task.
%   Our best suggestion for decoding method as in our setting is Beam search with beam size of 5.
\end{description}

\textbf{Novelty.} The key novelty of our work is as follows:
% \kla{Wannita updates this part too.}
% \kla{TODO later}
\revise{R3.5}{
\begin{itemize}
    \item \our~is the first to leverage the standard token type information for code completion with a variety of multi-task training techniques, which is different from existing work that leverages AST token type information.
    \item \our~is the first to extensively explore the sensitivity of the task weighting parameter and decoding methods in code completion.
    % \item \our~is the first for an empirical study on decoding methods for code completion
    \item \our~surpasses \revise{R3.15}{five} state-of-the-art code completion techniques in our setting and the CodeXGLUE Benchmark setting, achieving the highest performance in both token-level and line-level predictions.
    \item \our\footnote{https://huggingface.co/Wannita/PyCoder} is publicly available on HuggingFace together with the code dataset\footnote{https://huggingface.co/datasets/Wannita/PyCoder} and token type dataset\footnote{https://huggingface.co/datasets/Wannita/PyCoder-Type}. \revise{R3.17}{Our source code also available on GitHub\footnote{https://github.com/awsm-research/pycoder}.}
\end{itemize}
}

\textbf{Paper Organization.} 
The paper is organized as follows.
Section~\ref{sec:background} describes the background, motivation, and limitations of the state-of-the-art approaches.
Section~\ref{sec:approach} presents our \our~approach. 
Section~\ref{sec:experiment} describes the experimental setup and state-of-the-art baselines. 
Section~\ref{sec:results} presents the experimental results and discussions. 
Section~\ref{sec:discussion} discusses the results of our \our. 
Section 7 describes related work to code completion.
Section~\ref{sec:threats to validity} discloses the threats to validity.
Section~\ref{sec:conclusion} draws the conclusion.

% The model architectures on \gls{mtl} and \gls{ifn} are explained in session 3. The experimental part is in session 4 which includes dataset (4.1), model training (4.2), post-processing methods (4.3), and baselines information (4.4). The experimental results from 4 research questions are in session 5. Lastly is the discussion and related works in session 6 and 7.

% \begin{figure}
%     \centering
%     \includegraphics[width=\columnwidth]{figures/motivation2.png}
%     \caption{a) source code example for parsing. b) AST parsing result c) standard token types result}
%     \label{fig:ASTvsType}
% \end{figure}

% \begin{figure}
%     \centering
%     \includegraphics[width=\columnwidth]{figures/motivation.png}
%     \caption{a) source code example for parsing. b) AST parsing result c) standard token types result}
%     \label{fig:ASTfailvsType}
% \end{figure}

\section{Background and Motivation}\label{sec:background}

In this section, we discuss related work about automated code completion to situate the problems and present a motivating analysis.  

\subsection{Code Completion}
Code completion is a task to suggest the next code token from a given context. More formally, given a sequence of $m$ tokens $x_1 ... x_m$ as a context, code completion aims to predict the next $n$ tokens to complete a sentence $x_1 ... x_{m+n}$. 
The learning objective of a language model for code completion is to minimize a conditional probability distribution of the following function:
% standard language modelling objective that applied to code completion is a conditional probability distribution $P$ as follow:

\[ P(x_{1:m+n}) = \prod_{i=1}^{m+n} P(x_i|x_1 ... x_{i-1}) \]

% There are several approaches proposed for code completion. 
% Starting from the traditional tools by listing out all the possible attributes or methods that can invoke, heuristic rules~\cite{hou2010towards}, program history~\cite{robbes2008program}, and code examples~\cite{bruch2009learning}.

\textbf{Statistical language models.} Previously, several studies proposed code completion approaches using various types of techniques (e.g., heuristic, statistical, and deep learning).
Heuristic-based approaches aim to recommend source code based on rules~\cite{hou2010towards}, program history~\cite{robbes2008program}, and code examples~\cite{bruch2009learning}.
However, heuristic-based approaches are heavily based on rules and patterns that researchers need to develop, which is time-consuming and expensive.
Therefore, statistical language models have been proposed to automatically learn the naturalness of source code based on a probabilistic of the occurrence of source code.
For example, Hindle~\ea~\cite{10.5555/2337223.2337322, hindle2016naturalness} argued that source code is natural and repetitive (similar to natural language) and found that an n-gram approach can accurately predict the next code token based on a given context.
Raychev~\ea~\cite{raychev2016probabilistic} proposed TGEN, a probabilistic-based learning approach with decision tree structures.
However, the statistical language models are able to learn only the limited number of $n$ consecutive tokens (according to the $n$-gram algorithm), which does not reflect the nature of the source code that is usually long (i.e., long-term dependencies).

% Then there are statistical language models to capture the statistical patterns in source code using the occurrence probabilities of sequence of words such as n-gram~\cite{hindle2016naturalness} and decision tree~\cite{raychev2016probabilistic}. 

% Traditionally, code completion is formulated as a probabilistic language model (e.g., n-gram~\cite{hindle2016naturalness}) where the goal is to suggest the next probable code tokens based on a given context of the existing sequence of source code.
% To address various limitations of the traditional probabilistic language models (e.g., \kla{??}), 

% \kla{Gam will do the rest.}
% \kla{need to do this para with Wannita}

% together with the similar characteristic to natural language, code completion is shaped to Natural Language Processing (NLP) problem.
 % in Natural Language Processing (NLP)
 % The model is designed specially to solve the code completion Out-of-Vocabulary (OOV) problem.

\textbf{LSTM-based language models.} To address the limitation of the statistical language models, Long Short-Term Memory (LSTM)-based deep learning approaches are applied to the code completion task.
However, existing LSTM-based language models can only learn the semantic information of the source code, without considering its syntactic structure.
Thus, to ensure that the LSTM-based code completion models recognize the syntactic information, Abstract Syntax Tree (AST) is widely used by the previous work.
For example, Li~\ea~\cite{li2017code} proposed Pointer Mixture Networks, which is an LSTM-based architecture for predicting the AST node. 
Similarly, Svyatkovskiy~\ea~\cite{svyatkovskiy2019pythia} proposed Pythia, which is an LSTM-based approach that incorporates ASTs information through the Word2Vec embedding approach.
While such RNN-based and LSTM-based are able to handle longer sequences of source code than statistical language models, the approach remain inaccurate due to the sequential nature of source code processing, the limited ability to capture long-term dependencies, and the limited ability to recognize the importance of different code tokens.

% Additionally, together with a more complex model, the data such as ASTs information has also been introduced to train the model on syntactical structures of source codes.
% Nonetheless, the early state of deep learning architectures still consider the weight parameters of each token equally which in fact some source code tokens may be less or more important.
% can recognize longer sequences  by recursively inputting the previous output state to the current step. \kla{out of context}
% Next after code completion is shaped to \gls{nlp} problem. There are the coming of deep learning for code completion. The deep neural networks such as RNNs and LSTMs are applied.

\textbf{Transformer-based language models.} 
To address the limitations of LSTM-based language models, the Transformer architecture is introduced for the code completion task.
% Particularly, with the attention mechanism~\cite{vaswani2017attention} inside the Transformer architecture, the model is able to  differentiate the importance of code tokens, allowing the models to pay attention to only the important tokens, not the less important ones.
Generally, the development of Transformer-based language models consists of two steps: pre-training and fine-tuning.
Pre-training is a process to train a Transformer-based language model in a self-supervised manner (i.e., without labels), allowing the language models to self-understand given data by itself (i.e., natural language or programming languages).
Normally, the language models for code completion are trained using a Causal Language Model (CLM) (i.e., predicting the unknown token after a sequence of known tokens).
Once a language model is pre-trained, the model is then fine-tuned on a specific dataset (e.g., PY150~\cite{raychev2016probabilistic}) with the same learning objective as the pre-training process (i.e., CLM).
For example, Lu~\ea~\cite{lu2021codexglue} proposed CodeGPT-based models, which is based on a GPT-2 architecture~\cite{radford2019language} that is pre-trained on both Natural Language (NL) corpus (i.e., WebText) and/or Programming Language (PL) corpus (i.e., CodeSearchNet)---i.e., PL only for CodeGPT, and NL+PL for CodeGPT-adapt.

% from Microsoft Research proposed
To ensure that the Transformer-based language models recognize the syntactic structure of source code, Kim~\ea~proposed TravTrans~\cite{kim2021code}, a vanilla Transformer-based language model that incorporates ASTs information through different encoding styles.
Similarly, Wang~\ea~\cite{wang2021code} leverages AST information with a vanilla Transformer-based language model, but using a different AST encoding technique (i.e., by flattening the ASTs nodes).
However, these AST-based code completion approaches 
% \fix{in both Transformer-based and Multi-Task Learning based} 
also leverage AST information at the inference phase, which requires source code to be completed at the inference time so the AST information can be parsed and obtained from the source code. 
Therefore, in practice, source code is often incomplete and not compilable (e.g., syntax errors), making the existing AST-based code completion approaches not applicable in real-world scenarios.

\begin{figure}[h]
    \centering
    \includegraphics[width=\columnwidth]{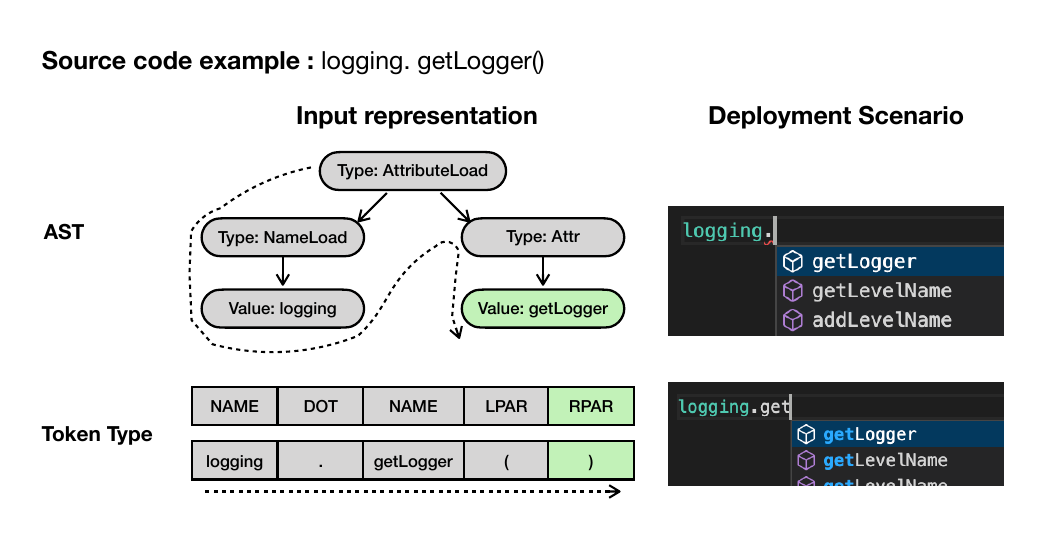}
    \caption{The comparison between AST and Token Type representations and the ideal deployment scenarios.}
    \label{fig:motivation}
\end{figure}

\subsection{A Motivating Example}

% In this section, we illustrate the limitation of an AST-based code completion approach through a motivating example followed by a motivating analysis.

% \kla{explain the analysis / parsable here / and present the findings. along with Figure 1 and new Figure on the results?}

% In this section, we conduct the analysis to prove the statement that ASTs information is not \emph{on-the-fly} information, i.e. in real-world scenario of code completion task, ASTs information is likely to be unable to extract which lead to the unseemliness of automate code completion.

% In this section, we explain our motivation of this paper.
% There are 2 main motivations: the AST node representation is not a natural order for code prediction, and the AST extraction require the complete and syntactically correct source code as inputs.
% Firstly, 

% \textbf{A Motivating Example.} 
Let's consider a code snippet \texttt{logging.getLogger()} as an example (see Figure~\ref{fig:motivation}).
\texttt{logging.} is the input code token, while \texttt{getLogger()} is the code token to be predicted.
Below, we illustrate two key limitations of the AST-based code completion approach by using TravTrans~\cite{kim2021code} as an example, which makes the existing AST-based code completion \emph{not able to predict next code tokens on-the-fly}.

\emph{First,} the learning objective of TravTrans does not reflect the natural order of typing source code sequences.
Since representing the source code as AST node sequence by traversing the AST, the order of the node sequence are inconsistent with the token sequence~\cite{liu2022unified}.
For example, at the learning phase, TravTrans~\cite{kim2021code} represents the input code tokens as a sequence of an AST node structurekl (i.e., [\texttt{AttributeLoad}, \texttt{NameLoad}, \texttt{logging}, \texttt{Attr}]) in order to predict the next AST node (i.e., [\texttt{getLogger}]).
However, this learning objective does not mimic the natural sequence of code tokens (i.e., [\texttt{logging}, \texttt{.}, \texttt{getLogger}, \texttt{(}, \texttt{)}]), meaning that the programming language-specific characters (e.g., dot [\texttt{.}] and parenthesis [\texttt{(},~\texttt{)}]) are currently ignored.
Therefore, in many cases at the deployment scenarios, such AST node information needs to be post-processed in order to successfully perform code completion in practice (e.g., add missing tokens [\texttt{(},~\texttt{)}], convert [\texttt{Attr}] to [\texttt{.}]).

\emph{Second,} in order to use AST information as an input, TravTrans~\cite{kim2021code} requires source code to be completed at the inference time so the AST information can be parsed and obtained from the source code. 
For example, in Figure~\ref{fig:motivation}, if developers type \texttt{logging.}, TravTrans can successfully recommend the next token (e.g., \texttt{getLogger)}).
However, source code is often incomplete and not compilable.
For example, in Figure~\ref{fig:motivation}, if developers type \texttt{logging.get}, TravTrans cannot correctly recommend the next token, due to the syntax errors during the AST parsing step.

\subsection{A Motivating Analysis}
\label{sec:motivation}

% \textbf{Motivating Analysis.} 
To demonstrate the significance of the problem of the AST-based code completion approaches, we perform a motivating analysis to investigate how often AST information could be provided at the inference phase, making AST-based code completion can be executed at the inference phase.

% \textbf{Approach.} 
Let's assume that a developer is typing a Python program character-by-character, we aim to analyze how often an AST parser can/cannot successfully parse a Python program at each character.
To do so, we select a statistical representative sample of 383 syntactically correct Python files from the PY150 dataset (with a confidence level of 95\% and a confidence interval of 5\%).\footnote{https://www.surveysystem.com/sscalc.htm}
Since we simulate the application of AST-based code completion at the character level, we execute a Python AST parser\footnote{https://docs.python.org/3/library/ast.html} at each character incrementally.
In total, we execute a Python AST parser for 1,263,296 times according to the total of 1,263,296 characters.
We find that 33.96\% of the executions can be successfully parsed, while 66.04\% of the executions fail to parse due to syntax errors.

\begin{table}[h]
    \centering
    \caption{The percentage of the successful/failed executions of the Python AST parser from the 1,263,296 executions.}
    \begin{tabular}{c|c}
        \textbf{AST Parsable?} & \textbf{Percentage} \\
        \hline
        Successful executions & 33.96\% \\
        Failed executions & 66.04\%
        % Success & 28.68\% \\
        % Syntax Error & 71.32\%
    \end{tabular}
    \label{tab:simulation}
\end{table}

\begin{tcolorbox}
\emph{\textbf{Finding:} For every two out of three characters that developers type, AST-based code completion cannot be performed at all due to the failed execution of the Python AST parser, limiting its ability to perform code completion on-the-fly at the inference time.
Since existing syntax-aware code completion is not on-the-fly and existing on-the-fly code completion is not syntax-aware, this paper aims to address these significant gaps by proposing a syntax-aware on-the-fly Python code completion approach.
}
\end{tcolorbox}

\section{Syntax-Aware On-the-Fly Code Completion}\label{sec:approach}

In this section, we present an overview of our syntax-aware on-the-fly Python code completion approach (\our).

Conceptually, \our~aims to generate source code at any time regardless of the completeness of the source code, while considering the syntactic and semantic information of the source code during the learning phase, but \emph{do not} require syntactic information during the inference phase.
To ensure that the learning process considers both semantic and syntactic information, we design our approach to focus on two prediction tasks, i.e., the code token prediction task and the token type prediction task.
In particular, we leverage a Multi-Task Training technique (MTT) to cooperatively learn both the code token prediction task (Task 1: Predict the next code token, considered as a Target Task) and the token type prediction task (Task 2: Predict its token type, considered as a Supporting Task).
For the type prediction task, we propose to leverage the standard Python token type information (e.g., String, Number, Name, Keyword), which is readily available and lightweight, instead of using the AST information~\cite{kim2021code, izadi2022codefill, li2017code, svyatkovskiy2019pythia, liu2020self, liu2022unified} where we found not available for the two-third of the executions (see our finding in Section~\ref{sec:motivation}), limiting its ability to perform on-the-fly code completion.
In contrast, our \our~\emph{does not} require syntactic information at the inference phase.
Thus, the completeness of the source code at the inference time is not required.

\begin{figure*}
    \centering
    \includegraphics[width=\textwidth]{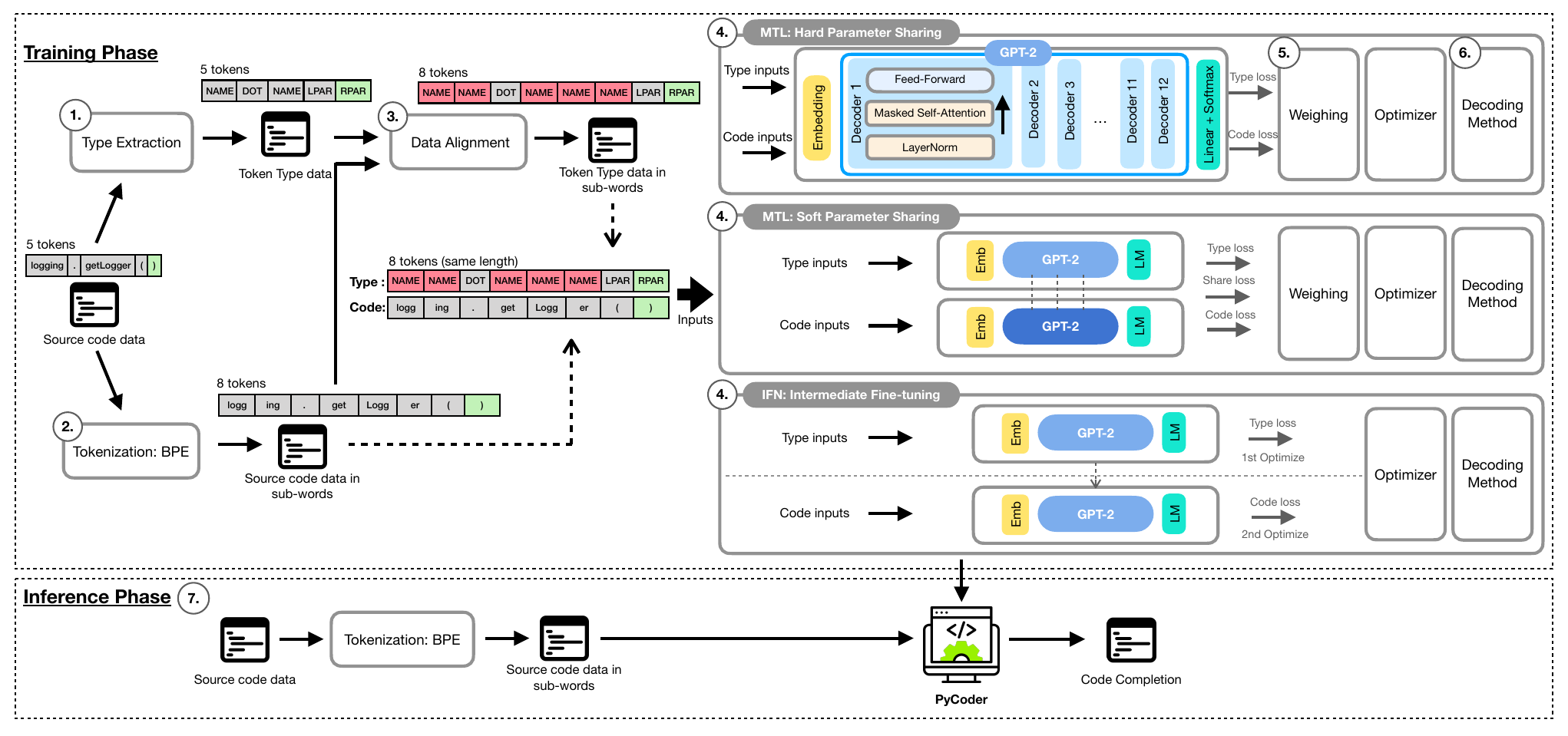}
    \caption{
    % \kla{An} \kla{o}verview of our \kla{syntactic-aware on-the-fly code completion} approach (\kla{SynComp})\kla{.} \kla{add a full stop in the caption everytime.} \kla{remove tag when reading this one.} \kla{This figure is nicely presented. Well done! Looks very neat.}
    An overview of our Syntax-Aware On-the-Fly Python Code Completion approach (PyCoder). 
    % \kla{can you use 1,2,3 instead of 3.1, 3.2, 3.3? Thank you.}
    }
    \label{fig:overview}
\end{figure*}

\textbf{Overview.} Figure~\ref{fig:overview} presents the overview of our \our, which consists of two phases: training and inference.
During the training phase, \our~performs 6 main steps:
Step~\circled{1} Type Extraction, to extract the token type information from source code;
Step~\circled{2} Tokenization, to perform subword tokenization on the source code;
Step~\circled{3} Data Alignment, to align the type information which is word level to the code information which is currently subword level;
Step~\circled{4} Multi-task Training Architecture with 3 training techniques: hard parameters sharing (MTL), soft parameters sharing (MTL), and intermediate fine-tuning (IFN);
then in Step~\circled{5} Hyperparameter Task Weighing and Step~\circled{6} Decoding Methods are the exploration steps to maximize the performance.
For the inference phase, we describe in Step~\circled{7} Code Generation step in the details of token-level prediction and line-level prediction.
% Below, we present the details of each step.

% \textbf{Problem Formulation.} 
% We perform the experiments on 2 kinds of completion level: token-level prediction and line-level prediction.

% MTL - Hard/Soft
% IFT - STILTs

% First, in \emph{(3.1) Data Collection} we extract the token type information from the source code. Then, in \emph{(3.2) Data Processing}, we tokenize the source code with \gls{bpe} algorithm and align the type information to the sub-words. Thus, the training phase consist of 2 tasks: source code prediction and token type prediction. In \emph{(3.3) Model Architectures}, we proposed 3 training techniques: hard parameters sharing MTL, soft parameters sharing MTL, and intermediate fine-tuning. For the inference phase, the model predict the task separately requiring no additional data for each task. Following is the details of our approach.

% We describe our approach in this session. Firstly, we elaborate how we collect the type dataset. Then since our model use BPE tokenizer, we align the type dataset to our source code to be in the same sequences. Lastly, we describe the models experiments which consist of 3 kinds: hard parameters sharing model, soft parameters sharing model, and intermediate fine-tuning model.

\subsection{(Step 1) Type Extraction}

Syntactic information can be represented in many forms, e.g., Abstract Syntax Tree (AST) which is widely used in the previous work, and Token Type information which remains largely unexplored.
In fact, both AST and token type information have their own advantages and disadvantages.
While AST provides a formal representation of syntactic information of source code, it requires syntactically correct source code in order to be successfully parsed by a Python AST parser.
Since our finding in Section~\ref{sec:motivation} shows that the Python AST parser failed to execute for every two out of three characters that developers type, the usage scenarios of the existing AST-based code completion approach are still limited in practice.

To address this challenge, we leverage a standard Python token type information, offering a more abstract representation of the syntactic structure of source code (e.g., Name, String, Number), which (1) is more lightweight, (2) follows the natural order of code sequences; and (3) can be successfully parsed at any times without requiring the complete and syntactically correct source code.
Generally, the standard Python token consists of two pieces of information i.e., (1) the token type, which provides syntactic meaning, and (2) the token value, which provides semantic meaning.
% Thus, each token is a substring that has semantic meaning in the grammar of the Python programming language.\footnote{https://www.asmeurer.com/brown-water-python/intro.html}
For example, given a \texttt{logging} token, the token type is \texttt{NAME} and its value is \texttt{logging}.
Since the token type information is not available in the existing code completion benchmark, we describe the steps to extract the type information below.

To extract type information, we use the \texttt{tokenizer} function provided by the standard Python tokenizer library\footnote{https://docs.python.org/3/library/tokenize.html} with an option \texttt{exact\_type} in order to extract the most fine-grained type for each token.
For the Python tokenizer (Python 3.7 version), there will be a total of 58 different types.
In particular, we focus on the 12 primary types of code tokens as follows: \texttt{<NAME>}, \texttt{<NUMBER>}, \texttt{<STRING>}, \texttt{<INDENT>}, \texttt{<DEDENT>}, \texttt{<ERRORTOKEN>}, \texttt{<ENDCODING>}, \texttt{<ENDMARKER>}, \texttt{<COMMENT>}, \texttt{<NL>}, \texttt{<NEWLINE>}, and \texttt{<OP>}, where the \texttt{<OP>} type consists of the remaining 46 operational types (e.g., operator. delimiter), such as \texttt{<LESS>}, \texttt{<GREATER>}, \texttt{<EQUAL>}, \texttt{<DOT>}.
Then, we perform the following pre-processing steps.

\begin{itemize}
    \item First, we discard the following three token types that will not be executed, i.e., \texttt{<ENCODING>} which describes the encoding of the Python file, \texttt{<ENDMARKER>} which describes the end position of the Python file, and \texttt{<COMMENT>} which describes the code comment of the Python file.
    \item Second, \texttt{<NAME>} provided by the Python tokenizer could be either identifier names (e.g., \texttt{logging}) or Python reserved names (e.g., \texttt{True}).
    Thus, a code completion approach may not be able to recognize the difference between the identifier names and the Python reserved names---which does not reflect the reality.
    To ensure that our code completion approach can recognize the difference between different types of names, we use the \texttt{keyword.iskeyword()} function\footnote{https://docs.python.org/3/library/keyword.html} in order to check and rename all of the Python reserved words which is originally extracted as  \texttt{<NAME>} to \texttt{<KEYWORDS>}. 
    \item Third, since the CodeXGLUE~\cite{lu2021codexglue} benchmark dataset treats any new line equally, we also convert  \texttt{<NEWLINE>} (a new line), \texttt{<NL>} (a new blank/comment line) as \texttt{<EOL>} (the end of line).
\end{itemize}

With this approach, the representation of the token types (i.e., each token has its own type) follows the natural order of source code, not the AST structure which addresses the limitations of the AST-based code completion approaches.
As shown in Figure~\ref{fig:overview},  \texttt{logging.getLogger()} will be tokenized as \texttt{[logging, ., getLogger, (, )]} with the following token types \texttt{[NAME, DOT, NAME, LPAR, RPAR]}.

\subsection{(Step 2) Tokenization}
% Generally, source code is a sequence of code tokens where each token has its own semantic meaning.
% \gam{} Therefore, 
Tokenization is an important step in automated code completion, aiming to split the source code into meaningful units.
There are three general levels of granularity, i.e., a word level, a subword level, and a character level.
While the word-level representation is the simplest tokenization approach, it may produce a massive vocabulary size. 
However, limiting the vocabulary size based on its frequency may cause an Out-of-Vocabulary words (OOV) problem.
While the character-level representation can diminish the OOV problem with the limited vocabulary size (e.g., English characters), models may not be able to handle an excessively long sequence of source code (i.e., each character has its own vector).
Instead, we use sub-word tokenization with the Byte-Pair Encoding (BPE) algorithm~\cite{sennrich2015neural}, as prior studies found that BPE can substantially reduce the vocabulary size~\cite{fu2022gpt2sp, karampatsis2020big}, while being able to generate new identifiers that never appear in the dataset~\cite{thongtanunam2022autotransform}.
First, BPE splits source code into characters.
Then, BPE iteratively merges the characters into subwords based on the frequency of the occurrences to create the vocabulary until the desired size.
In this paper, we use the CodeGPT tokenizer, which has a vocabulary size of 50,000 subwords.
To ensure that the CodeGPT tokenizer can recognize the token types, we represent the token types in the bracket parenthesis form $\langle...\rangle$, which are included in the special token vocabulary for the BPE tokenizer to avoid any subword tokenization on these token types.

\subsection{(Step 3) Data Alignment}
Data alignment is an important step to ensure that the sequence of code tokens and their corresponding token types are correctly matched and aligned.
With the use of BPE, some words may be tokenized as subwords, while their type is not tokenized into the subword level, making the sequence of code tokens and the corresponding token types not correctly matched.
For example, as shown in Figure~\ref{fig:overview}, BPE splits \texttt{logging} into \texttt{[logg, ing]} with a single corresponding \texttt{<NAME>} token type.
To address this problem, we repeat the token type for any word that is split by BPE.
Therefore, in Figure~\ref{fig:overview}, the token type \texttt{<NAME>} is repeated twice in order to match the subword-level code sequence of \texttt{[logg, ing]}.
This data alignment step will produce a sequence of code tokens and their corresponding token types with the same length, which is ready to be fed into our code completion approach to learn both syntactic and semantic meanings of source code.

\subsection{(Step 4) Multi-Task Training Architectures}
\label{sec:approach-arch}

% On the other hand, training multiple single models on multiple tasks is also possible, but predictions are competing and the multiple sources of information are not jointly learned together, leading to inaccurate predictions.

Our \our~leverages a Multi-Task Training (MTT) paradigm, which is a set of techniques designed to learn multiple tasks, allowing the model to capture multiple sources of information.
Traditionally, deep learning is designed for one single learning objective (e.g., only predicting the next code token), limiting its ability to capture other important and useful sources of information (e.g., syntactic information of source code).
Instead of training a model with one single learning objective, the MTT paradigm aims to provide a generalist model with multiple learning objectives, providing a more robust vector representation.
For our \our~approach, we design the target task to predict the next token, while the supporting task (aka. an auxiliary task or additional related non-target task) is to predict the token type.
In addition, we build three variants of \our, with three different MTT techniques, according to two learning styles~\cite{phang2018sentence} as follows.
% (i.e., \our-Hard, \our-Soft, and \our-IFN)

\subsubsection{Multi-Task Learning (MTL)}

\revise{R1.2}{Multi-Task Learning (MTL) learns multiple tasks simultaneously instead of learning them separately.}
Normally, during the learning process, the model aims to optimize a loss function for one single learning objective.
With the MTL approaches, multiple loss functions are optimized together during the learning process, allowing the MTL-based model to simultaneously learn against multiple objectives and share the knowledge understanding from multiple related sources.
In this paper, we consider two main MTL approaches for Multi-Task Learning (MTL)~\cite{ruder2017overview}, i.e., Hard Parameter Sharing (\our-Hard) and Soft Parameter Sharing (\our-Soft).

For \emph{Hard Parameter Sharing}, the key principle is to train a code completion model against two learning objectives, where the loss functions of the two learning objectives ($L_{code}$ and $L_{type}$) are optimized together within the same model.
Formally, the \our-Hard model aims to minimize the following loss function: 
\begin{equation}
    \label{eq:1} 
    \begin{aligned}
    \medmath{L_{Hard} = \argmin_{\omega}(L_{code}(d_{code}, \omega) + L_{type}(d_{type}, \omega))}
    % loss_{hardShare} = codeLoss + typeLoss
    \end{aligned}
\end{equation} 

\noindent , where $d_{code}, d_{type}$ denotes the code token dataset and the token type dataset, respectively, and $\omega$ denotes a model's parameters.
With Hard Parameter Sharing, the weights and model parameters are shared between tasks, allowing the model to explicitly learn the input representations between tasks (i.e., code and type vectors) that are closely related.

% via the shared layers which may help enhance the effectiveness of the target task's results.
% After the shared layers, usually there are separated task-specific output layers for each task.
% However for our work, the tasks have similar outputs i.e. tokens from the vocabulary.
% Therefore, we train both target task and supporting task in the same model simultaneously without separated output layer.

% SynComp hard parameter sharing uses one GPT-2 model with two losses: code prediction loss ($L_{code}$) and type prediction loss ($L_{type}$).
% The model minimize the summation of losses shown in Equation~\ref{eq:1}.

% For hard parameter sharing, the tasks will share the same model backbone which mean the weights and parameter of the model being shared together.
% Only the task-specific output layer that might be different.
% A hard parameter sharing model is one of the most popular multi-task learning techniques.

% \kla{wannita will write soft....}

% \subsubsection{MTL: Soft Parameter Sharing Model}
% For soft parameter sharing, each task will have its own model backbone, i.e. separate weights and parameters.
% However, the models are loosely connected by the constraint which try to minimize the distance of the model parameters' difference.
% This is to regularize the model parameters to be similar.
For \emph{Soft Parameter Sharing}, the key principle is similar to Hard Parameter Sharing where the goal is to train a code completion model with two learning objectives.
However, instead of training a model against two tasks like the Hard Parameter Sharing model, the Soft Parameter Sharing is designed to train two individual models for each task ($L_{code}$ and $L_{type}$), allowing each model to learn separately for each task.
Therefore, each learning objective has an individual model (i.e., separated weights and parameters between the learning objectives).
% to learn and predict outputs; 
To allow the model to share the knowledge between tasks (i.e., to learn the similarities between the related parameters), a shared loss function is also used, 
\revise{R3.14}{which is computed from Euclidean norm~\cite{golub2013matrix}} as follows:

\begin{equation}
    \label{eq:norm}
    L_{sharing}(\omega_1, \omega_2) = 
    % ||W||_F = 
    \sqrt{\sum_{i=1}^{I}\sum_{j=1}^{J}|\omega_{1(i,j)}-\omega_{2(i,j)}|^2}
\end{equation}

% Nonetheless, the models are loosely connected by the constraint to encourage similarities between related parameters.

% The constraint ($L_{sharing}$) which is used to optimize the difference of distance between the models parameters' applies with Frobenius norm:

\noindent , where $\omega_n$ denotes the model parameters of the  learning objective $n$. Finally, the \our-Soft model aims to minimize the following loss function:

\begin{equation}
\label{eq:2}
\begin{aligned}
    L_{Soft} = \argmin_{\omega_1, \omega_2}
    &( L_{sharing}(\omega_1, \omega_2)\\ 
    &+ L_{code}(d_{code}, \omega_2)\\
    &+ L_{type}(d_{type}, \omega_1))
\end{aligned}
\end{equation}

With Soft Parameter Sharing, each learning objective has its own model parameters and weights, allowing the models to implicitly learn the input representations that might have more connection to a specific task.

% SynComp soft parameter sharing uses two GPT-2 models with three losses: code prediction loss, type prediction loss, and sharing loss. The model minimize the summation of losses shown in Equation~\ref{eq:2}.

\subsubsection{Intermediate Fine-Tuning (IFT)}

\emph{Intermediate Fine-Tuning (IFT)}~\cite{phang2018sentence} adapts a transfer learning concept (i.e., pre-training then fine-tuning) where the goal is to learn multiple tasks sequentially. First, the model is fine-tuned on the supporting task (token type prediction) followed by the target task (code token prediction), respectively.
Thus, the fine-tuned step on the supporting task can be considered the second stage of the model pre-training.
Therefore, the Intermediate Fine-Tuning (IFT) model (\our-IFT) is first trained based on an intermediate self-supervised task (token type prediction), then trained on the target task (code token prediction), allowing the model to gain knowledge on the token type prior to predicting the next code tokens.

\subsection*{GPT-2 Model Architecture}

Among the three variants of the MTT techniques (i.e., \our-Hard, \our-Soft, and \our-IFT), we use the GPT-2 architecture as a base model.
GPT-2~\cite{radford2019language} is a decoder-only Transformer model.
The GPT-2 architecture for code completion consists of three main components: the embedding layer, the decoder block, and the language model head. 
First, the embedding layer embeds the input tokens into vectors with positional encoding, allowing the model to learn the semantic meaning and the position of each code token.
Then, the embedding vectors are fed into the decoder block which contains decoder layers.
Each decoder layer includes masked self-attention layers, feed-forward neural network layers, and normalization layers.
%  It basically always scores the future tokens as 0 so the model can’t peak to future words
The masked self-attention layer indicates which tokens to focus on, while the masking approach prevents the attention mechanism~\cite{vaswani2017attention} to see the unseen tokens in the future.
% \kla{wannita will do the rest}.
% Feedforward neural nets are complex network made up an input layer that accepts information, hidden layers that capture the hidden correlations between each data point, and an output layer which transmit information.
The feed-forward neural network layer is a sophisticated network with hidden nodes to capture the related information between each data point.
% Layer normalization (LayerNorm) is a technique to normalize the distributions of intermediate layers. It enables smoother gradients, faster training, and better generalization accuracy.
The normalization layer makes the learning process more effective by enabling smoother gradients and generalized accuracy.
% The linear layer takes the output of the last decoder block and converts it to a vector whose dimensions are vocabulary size by 1. In short, it takes a lot of inputs and produces a list where each spot represents a token. The higher the number in the spot the better the chance that that token is the best pick. Softmax converts the output of the linear layer to a probability distribution
After $L$ layers of decoder, an output of the last layer is fed to the language model head, i.e., a linear layer, which converts the output to a vector whose dimensions are the same as the vocabulary size.
Lastly, the vector is converted to a probability distribution by the softmax activation function.
Formally, to predict the next token $x_t$ based on a given input sequence, GPT-2 can be represented as follows:

\begin{equation}
\begin{aligned}
    \label{eq:transformer}
    h_0 &= W_e \cdot C + W_p \\
    % \label{eq:transformer2}
    h_l &= decoder\_layer(h_{l-1}), \forall l \in [1,L] \\
    % \label{eq:transformer3}
    P(x_t) &= y_t = softmax(h_n \cdot W^T_e), t \in [0, N]
\end{aligned}
\end{equation}

\noindent, where $W_e$ is the tokens embedding matrix, $C$ denotes the context vector of tokens, $W_p$ is the position embedding matrix, $L$ is a number of decoder layers, and $N$ is the length of the sequence.
We follow the traditional language models by maximizing the log-likelihood of:

\begin{equation}
    \label{eq:log-likelihood}
    L(x_t) = \sum_i{\log P(x_i|x_1...x_{i-1}, \omega)}
\end{equation}

\noindent, where $\omega$ is the model parameters that are learned during the optimization process.
Particularly, \our~uses the pre-train CodeGPT~\cite{lu2021codexglue} that is pre-trained on the CodeSearchNet dataset~\cite{husain2019codesearchnet} as a starting checkpoint.

\subsection{(Step 5) Hyperparameter Task Weighting}
\label{sec:approach-weight}

Since our \our~leverages MTL training techniques to learn multiple different tasks simultaneously, some tasks may have a higher influence than others, which later may produce an unsatisfactory accuracy for the other tasks (called a conflicting gradient problem).
To prevent such conflicting gradients between tasks, it is important to find the most optimal task weights by minimizing the loss.
Therefore, we optimize the hyperparameters ($\alpha_i$) to adjust the task weights to find optimal task weights for our architecture.
Specifically, we aim to minimize the loss of the code prediction task along with the type prediction task using the following loss function.

\begin{equation}
    \label{eq:rq3}
    L_{MTL} = \argmin_{\omega}(\sum_{i}\alpha_i \cdot L_{i}(d, \omega))
    % L_{MTL} = \min_{\omega}(\alpha * L_{code}(\omega) + (1 - \alpha) * L_{type}(\omega))
    % loss = \alpha * codeLoss + (1 - \alpha) * typeLoss
\end{equation} 

% \our~performs a static weighted linear sum of losses which fixed the task's weights throughout the training phase.
% In Equation.~\ref{eq:rq3} presents SynComp's task weighing formula where $\alpha$ is the hyperparameter weight. 

\subsection{(Step 6) Decoding Methods}
\label{sec:approach-decoding}

Decoding is a method to select the next token from the potential vocabulary when generating a sequence.
Although selecting only the highest probable token is suitable for a single step, it might be sub-optimal for the sequence.
Since the search space of the next tokens is large, different decoding methods will have different mechanisms, providing different predictions of the next tokens.
Thus, the selection of the decoding methods may have an impact on the overall performance of our ~\our.
In the code completion literature, we found that Beam Search is one of the most commonly used decoding methods.
However, Holtzman~\ea~\cite{holtzman2019curious} found that there exist other decoding methods that are widely used in the NLP area, yet remain largely explored in the code completion literature.
Thus, we aim to experiment with the six following decoding methods.

\begin{itemize}
    \item \textbf{Greedy} is a method to select the maximum probable vocabulary to be the next tokens.
    This method assumes that the model already outputs the best probability in every timestep.
    % it generates all possible tokens in the vocabulary list; then, it will choose top B candidates that have the most probability. Those B candidates will move to the next time step, and the process repeats. In the end, there will only be B candidates. The search space is only (10,000)*B.
%     Although selecting only the highest
% probability token is suitable for a specific time step, it
% might be a sub-optimal for a sequence. 
    
    \item \textbf{Beam Search} applies a search algorithm to generate all possible tokens in the vocabulary; then, it selects the top $b$ (i.e., beam size) probable tokens to continue.
    The Beam Search method is one of the most commonly used decoding methods in text generation tasks~\cite{li-etal-2016-deep, wiseman-etal-2017-challenges}.
    % However, Beam Search may not always achieve optimal results, since it does not consider the whole vocabulary, but instead only the top $b$ (i.e., beam size) probable tokens.
    % Nevertheless, Beam Search still performs faster than an exhaustive search.
    
    % ; however, it balances between the performance and the computational time.
    % The method's time complexity is equals to $O(b*V)$ where $V$ is the vocabulary size.
    % It . 
    
    % recommend the most probable $b$ next tokens according to a defined $b$ beam size threshold.

    % by expanding the graph in a limited set called beam~size~($b$).
    % In each timestep, the method generates all possible tokens in the vocabulary; then, it select the top $B$ probability tokens to continue.

    \item \textbf{Sampling} is a method to randomly select the next token from the actual probability distribution assigned by the model.
    Different from Greedy and Beam search methods which in some cases may recommend only the same probable next tokens at different timesteps, the sampling method may recommend different next tokens at different timesteps (i.e., non-deterministic).
    % We put the sampling method in this study as the baseline for other decoding methods. All of the methods that required sampling are set the seed value to the same number.
    
    % Temperature is used to increase the probability of probable tokens while reducing the one that is not. Usually, the range is 0 < temp ≤ 1. Note that when temp=1, there is no effect.
    \item \textbf{Sampling with Temperature} applies a temperature parameter to shape the probability distribution~\cite{ackley1985learning}, which is different from the original sampling method where the randomness is arbitrary.
    The temperature is used to increase the probability of the most probable next tokens, while decreasing the probability of the others.
    We note that the probability of the least probable next tokens is only decreased, but they are not removed from the recommendation.
    The range of the temperature value is usually at $0 < temp \le 1$, where $temp = 1$ is a normal sampling.

    % Additional to normal sampling which could be arbitrary, sampling with temperature method 
    \item \textbf{Top-K Sampling} aims to truncate the probability distribution by choosing the top-$k$ probable next tokens from the vocabulary, then, re-scale the distribution and perform sampling based on the new distribution.
    This method ensures that the less probable next tokens will not be generated, while only the top-$k$ probable next tokens are only considered during the sampling process.
    
    \item \textbf{Top-P Sampling (Nucleus Sampling)} is similar to the Top-k sampling method where the Top-P sampling method also truncates the probability distribution, but with different criteria. 
    Top-P sampling prunes the distribution by the cumulative probability of the current step $\ge p$~\cite{holtzman2019curious}; then, re-scale and perform sampling.
    Formally, given the probability P, we can define the smallest summation of the probability as $V_p$ in
    \begin{equation}
        \label{eq:top-p}
        \sum_{x\in V_p} P(x|x_{1:i-1}) \ge p
    \end{equation}
    The benefit of this method is that it can dynamically adjust the number of $k$ depending on the certainty of the model.
    If the model is very certain on some tokens, the search space is small, and vice versa.
\end{itemize}

%  the decoding method defines the way the system handles its search space over potential output utterances when generating a sequence
% The result performance in code completion may not depend only on the model, but also a method to select the output tokens from the probability distribution, i.e. a decoding method.
% The decoding method defines the way to handle the search space over potential output tokens when generating a sequence.
% Although selecting only the highest probability token is suitable for a specific time step, it might be a sub-optimal for a sequence.
% Therefore, various alternative way for decoding methods is proposed in NLP field; however, there is still a limited exploration in code completion which is similar to text generation field.
% Thus, we select 6 following decoding methods to study in this work in order to seek for the best use of our architecture.

% Even though the model leverages the token type prediction as the supporting task during the training process, at the inference phase, the model will perform only the code token prediction.
% Thus, the token type information is not required at the inference phase, allowing our \our~to perform on-the-fly code completion.

\subsection{(Step 7) Code Completion}
\our~performs predictions at two granularity levels, i.e., at the token level and at the line level.

\textbf{Token-level code completion} is a process to predict the next token (the right side), given the prior code tokens as a context (the left side).

\textbf{Line-level code completion} is similar to the token-level prediction, but the model aims to predict the next tokens until completing the whole line of code (i.e., not just only one single next token).
For the line-level prediction, we leverage the same model used for the token-level code completion task to iteratively generate the next token, where the newly generated token is used as a context for the next step of the prediction.
This process is repeated iteratively until the model generates a $\langle EOL \rangle$ token, or until it reaches a certain $n$ threshold ($n=100$, following the CodeXGlue~\cite{lu2021codexglue}).

\section{Experimental Setup}\label{sec:experiment}

In this section, we present the goal of our experiment, along with the research questions, followed by the experimental setup in detail.
% We describe our python (4.1) dataset and our (4.2) pre-processing methods which are adjusted from CodeXGLUE.
% Then, we describe the details of our implementation on (4.3) model training and (4.4) evaluation measurement.
% Lastly, we explain our four state-of-the-art (4.5) baselines.
% We present: our (4.1) Dataset, which the models train and test on PY150 dataset; our (4.2) Pre-processing method, which is adjusted from CodeXGLUE.; our (4.3) Model Training, about the parameters; (4.4) Evaluation Measurement; and (4.5) Baselines, which is 4 State-of-the-art.

% We present our (4.1) Dataset, (4.2) Pre-processing Methods, (4.3) Model Training, (4.4) Evaluation Measurement, and lastly (4.5) 4 State-of-the-art Baselines.

\subsection{Goal and Research Questions}

The goal of this paper is to empirically evaluate our \our~and compare with the state-of-the-art approaches according to the token-level and line-level code completion tasks and to provide a better understanding of the impact of the components of our \our. 
To achieve this goal, we present the motivation and the research questions below.

\begin{description}
    \item \textbf{RQ1) \rqone} \\
    \textbf{Motivation.} 
    As motivated earlier, existing syntax-aware code completions are not on-the-fly, while existing on-the-fly code completions are not syntax-aware.
    To address this important gap, we introduce \our~(a syntax-aware on-the-fly code completion).
    Thus, we formulate this RQ to investigate how well our \our~perform when compared to the state-of-the-art approaches for both token-level and line-level code completion tasks based on the CodeXGlue Benchmark.
    % In this work, \our~addresses the limitations of state-of-the-art approaches as discussed in section~2.
    % In particular, 
    % instead of leverage either AST information or only source code like state-of-the-art approaches, 
    % the goal of \our~is to perform the code completion with lightweight syntax-aware information while remaining the on-the-fly feature.
    % Thus, to evaluate that our syntactic information (i.e., token type) benefit the model,
    % we formulate this RQ to compare our \our~to CodeXGLUE benchmark and four state-of-the-art approaches which include both AST and non-AST approaches.
    % and also CodeXGLUE benchmark
    \item \textbf{RQ2) \rqtwo} \\
    \textbf{Motivation.} 
    There exist various training strategies for multi-task learning used in code completion.
    For example, Liu~\ea~\cite{liu2022unified} found that hard parameter sharing performs best, while Izadi~\ea~\cite{izadi2022codefill} found that soft parameter sharing performs better than hard parameter sharing for code completion.
    % In addition, an alternate approach widely used in NLP (i.e., Intermediate Fine-Tuning) has not been explored.
    This contradictory finding motivates us to investigate the impact of training strategies on the performance of \our.
    % leaapproach; however, there are various ways available to alternate the use of supporting task (e.g., IFN and different ways to manipulate data for MTL).
    % In NLP field, Weller~\ea~\cite{weller2022use} study on these different training techniques; yet, little knowledge is known about code completion.
    % Therefore, we formulate this RQ to investigate the impact of training techniques on our \our~code completion.
    \item \textbf{RQ3) \rqthree} \\
    \textbf{Motivation.}
    Our \our~relies on two prediction tasks, i.e., code prediction and token type prediction tasks. 
    It could be possible that these two tasks may be conflicting with each other or one task has a higher influence than the other task during the learning process. 
    Thus, prior studies ~\cite{vandenhende2021multi,sener2018multi} raised concerns that the conflicting issue (aka. conflicting gradient) may degrade the performance of multi-task learning.
    Therefore, task weighting parameters are used to weigh the importance of each task to achieve optimal accuracy.
    However, \our~may be sensitive to the task weighting parameters.
    Thus, we set out this RQ to investigate the impact of the task weighting parameters on the performance of \our.

    % ensure that different tasks that are learned simultaneously have equal importance during the learning process.
    % This w

    % are an important mechanism used in multi-task learning to weigh the importance of each task during the learning process.

    % Prior studies~\cite{vandenhende2021multi} raised concerns that different task weighting may have an impact on the performance of the MTL models. 
    
    % This means that when training multiple tasks simultaneously, one task that is weighted more than another may have a stronger influence on the predictions than the other task, which could result in sub-optimal performance.

    % \kla{I don't get this one --- will discuss with Wannita later}
    % While MTL approach achieves success in many deep learning application, many studies mention about task weighing problem~\cite{vandenhende2021multi}.
    % In other words, when training multiple tasks simultaneously, the model may face a problem of one task dominantly influence others resulting in the sub-optimal performance.
    % To avoid this problem, the importance of task's weights adjustment has been highlighted~\cite{sener2018multi}.
    % Thus, we formulate this RQ to explore the impact of the task weighing parameters in our \our.
    \item \textbf{RQ4) \rqfour} \\
    \textbf{Motivation.}
    Decoding methods are an important component of code completion used to generate the next probably code tokens.
    Recently, only a few methods are used for code completion (e.g., Beam Search, Greedy)~\cite{svyatkovskiy2020intellicode, kim2021code, izadi2022codefill, li2017code}.
    However, there are other decoding methods that have been used for text generation in the natural language processing field, yet have never been explored in software engineering.
    Thus, there is a lack of understanding of whether decoding methods widely used in code completion are the best.
    % Many decoding methods have been proposed for , and many of them are widely adopted in software engineering.
    % Beam Search seems to be 
    % Holztman~\ea~\cite{holtzman2019curious} study on the impact of different decoding methods to generate English text; however, little knowledge has been explore on decoding methods in code completion field.
    % Specifically, most of the time only Beam search method or Greedy method is selected off-the-self~\cite{svyatkovskiy2020intellicode, kim2021code, izadi2022codefill, li2017code}.
    % Thus, we formulate this RQ to explore the impact of the different decoding methods to our \our~code completion.
\end{description}

\subsection{Dataset}\label{dataset}

We use the ETH PY150 python dataset (the standard code completion benchmark) provided by Raychev~\ea~\cite{raychev2016probabilistic} to ensure a fair comparison with prior studies ~\cite{kim2021code, izadi2022codefill, li2017code, wang2020towards}.
The dataset is collected from open-source software projects in GitHub repositories with non-viral licenses (e.g. MIT, Apache, and BSD)---a license that an owner gives permission for freely use under specific terms; thus mitigating potential licensing issues.
Note that this dataset is also used in Microsoft's CodeXGLUE benchmark~\cite{lu2021codexglue}---a worldwide competition for the AI4Code area.
As any duplicated codes have been removed by Raychev~\ea~\cite{raychev2016probabilistic}, arriving at a total of 150,000 Python files, we confirm that there is no \revise{R3.9}{exact} code duplication between the training set and the testing set, thus mitigating several potential biases like code duplication in our experiment.
% \kla{need to double check} After the authors remove duplicated files and filter only up to 30K AST nodes, the dataset contains 150,000 python source code files in total.
Following CodeXGLUE, for token-level predictions, the dataset is split into 95,000 files for the training set, 5,000 files for the validation set, and 50,000 files for the testing set, with the number of tokens of 72.1M, 4.4M, and 37.3M, respectively.
For line-level predictions, it's a common practice to reuse the same model trained for token-level predictions. 
Thus, only a testing set is required, but a training set and a validation set are not required.
Therefore, we use the 10,000 Python files provided by CodeXGLUE~\cite{lu2021codexglue} as a testing set for line-level predictions.

\subsection{Pre-processing Methods}
Sensitive data information (e.g., name, number, credential, IP address) could appear in the source code.
To avoid the models unnecessarily paying attention to this information, we mask these sensitive data by creating a placeholder for any string and numeric literals in the source code.
Particularly, following CodeXGLUE~\cite{lu2021codexglue}, we first identify tokens based on their STRING and NUMBER types.
Then, in the top-200 most frequent strings and the top-30 most frequent numeric literals, we replace the string with $\langle$STR\_LIT:\textit{value}$\rangle$ 
 and replace the number with $\langle$NUM\_LIT:\textit{value}$\rangle$.
\revise{R3.10}{Note that we use similar frequent numbers to CodeXGLUE~\cite{lu2021codexglue}}.
The rest of the uncommon literals are masked by $\langle$STR\_LIT$\rangle$ or $\langle$NUM\_LIT$\rangle$.
Finally, these placeholders are also added to the special tokens of the tokenizer, avoiding any subword tokenization for these special tokens.

% the most 200 frequent

% We use CodeXGLUE~\cite{lu2021codexglue} methods to mask the literals of t with placeholders.
% Specifically, the most 200 frequent string and the most 30 frequent numerical literals are normalized in $\langle$STR\_LIT:\textit{value}$\rangle$ or $\langle$NUM\_LIT:\textit{value}$\rangle$ formats.
% The rest uncommon literals are masked by $\langle$STR\_LIT$\rangle$ or $\langle$NUM\_LIT$\rangle$.
% These placeholders are also added to the special tokens in the tokenizer; thus, they are not separated in the tokenization step (section 3.2).

% In this step, the string and numeric literals in source code dataset are normalized.
% The purposes are to mask the sensitive data that sometimes developers leave in the code  and for better training as we do not encourage the model to emphasize on different literals when predicting. 
% We use CodeXGLUE~\cite{lu2021codexglue} methods to mask the literals of token type STRING and NUMBER with placeholders.
% Specifically, the most 200 frequent string and the most 30 frequent numerical literals are normalized in $\langle$STR\_LIT:\textit{value}$\rangle$ or $\langle$NUM\_LIT:\textit{value}$\rangle$ formats.
% The rest uncommon literals are masked by $\langle$STR\_LIT$\rangle$ or $\langle$NUM\_LIT$\rangle$.
% These placeholders are also added to the special tokens in the tokenizer; thus, they are not separated in the tokenization step (section 3.2).

In addition, we preserve the original indentation of the source code that is ignored by CodeXGLUE's pre-processing step.
Indentation plays an important role as part of the Python syntax grammars, as it is used to indicate a group of statements that belongs to a particular code block, assisting a Python interpreter to decide the execution of the next statement.
To do so, for any positions of the indentation, we use $\langle$INDENT$\rangle$ and $\langle$DEDENT$\rangle$ special tokens.
$\langle$INDENT$\rangle$ denotes the indentation, which appears once at the beginning of a code block, \emph{not once per line}, while  $\langle$DEDENT$\rangle$ denotes the dedentation at the end of the code block.
% Thus, every $\langle$INDENT$\rangle$ token is matched by a corresponding $\langle$DEDENT$\rangle$ token.
% Similarly, these 
% We discard the indentation only when training the model for CodeXGLUE benchmark in RQ1 (section 5).

% assists on deciding which statement to execute next.

% Because the indentation is important in python language syntax as it assigns a group of statements to a particular block of code and assists on deciding which statement to execute next.
% We store the position of the indentation using $\langle$INDENT$\rangle$ and $\langle$DEDENT$\rangle$ tokens.
% While the $\langle$INDENT$\rangle$ token is presented once at the beginning per block of code, not once per line; the $\langle$DEDENT$\rangle$ token denotes the dedentation or the end of the block.
% Thus, every $\langle$INDENT$\rangle$ token is matched by a corresponding $\langle$DEDENT$\rangle$ token.
% We discard the indentation only when training the model for CodeXGLUE benchmark in RQ1 (section 5).

% \begin{table}[]
%     \centering
%     \begin{tabular}{l|c|c|c}
%         Dataset & Success & Indentation Error & Syntax Error \\
%         \hline
%         CodeXGLUE & 11.51 & 87.84 & 0.65 \\ 
%         Our & 88.07 & 0.00 & 11.93 \\
%     \end{tabular}
%     \caption{Compare the parsing rate between datasets.}
%     \label{tab:parsing rate}
% \end{table}

\subsection{Model Training}

% model backbone & hyperparameter -> move to model
% We use GPT-2, a decoding model consisting of 12 layers of transformers, as the based model for all our architectures.
% The warm-up checkpoint is the pre-train checkpoint from CodeGPT~\cite{lu2021codexglue} which is trained on 1.1M python function from CodeSearchNet dataset~\cite{husain2019codesearchnet}.
% Then we fine-tuned both our models and baselines on PY150 dataset. 

% tokenizer -> move to tokenization
% We adapt our tokenizer from CodeGPT. The vocabulary size for BPE method is set to 50,000.
% Additionally, the placeholders for types, and for sensitive data in strings and numbers are included in as the \emph{special tokens}. Thus, such placeholders are not split into sub-word.
% Overall, the final size of TypeComp's vocabulary is 50,288.

% For the tokenizer, we also use tokenizer from CodeGPT. The tokenizer is \gls{bpe} train for 50,000 token vocabs. The placeholder for sensitive data in strings and numbers are included in as the special tokens. Additionally, we also add all token types as the special tokens. Therefore the final size of vocabulary is 50288.

% training
We use PyTorch\footnote{https://pytorch.org}~\cite{paszke2019pytorch} and HuggingFace\footnote{https://huggingface.co}~\cite{wolf2019huggingface} libraries for the implementation of our GPT-2 based model with the pre-trained checkpoint of CodeGPT.
The base model is the default GPT-2 small configuration~\cite{radford2019language},   consisting of 12 layers of Transformer decoders, 12 attention heads, $n\_position$~=~1024, $n\_ctx$~=~1024, and $n\_embd$~=~768.
We train our models for 200,000 steps with an Adam optimizer~\cite{kingma2014adam}. 
The hyperparmeters setting is shown in Table~\ref{tab:1}.
We do not fine-tune the hyperparameters due to limited resources.
Therefore, our results could serve as a lower bound, but the optimization may improve the accuracy of our model.
Overall we train 12 variants of \our~(3 multi-task training techniques + 9 task weighing parameters) for a total of more than 850 training hours.
For the baseline, we use all the best hyperparameters described in their papers.
Our experiments is run on one NVIDIA GeForce RTX 3090 GPU with 24 GB memory, an Intel(R) Core(TM) i9-9980XE CPU @ 3.00GHz with 36 core processors, and 64G RAM.

% \kla{may be mention model training time and inference time, how many models are built, which is an approx of X GPU hours (~? days) for a commodity GPU}.

% n_positions=1024,
% n_ctx=1024,
% n_embd=768,
% n_layer=12,
% n_head=12,

% seqeunce?
% For RQ2 we compare our 3 architecture models without tasks weighing parameters.
% Then we use the best model regarding to the results from RQ2 to fine-tune tasks weighing parameters in RQ3.
% Combine results of RQ2 and RQ3, we use the best model and task weighing parameters to test for decoding methods in RQ4.
% Lastly, TypeComp, which has the best model architecture, tasks weighing parameters, and decoding method, is used to compete the state-of-the-art baselines in RQ1.

\begin{table}[]
    \centering
    \begin{tabular}{c|c}
        Hyperparameter & Value \\
        \hline
        Learning rate & 8e-5 \\
        Weight decay & 0.01 \\ 
        Batch size & 2 \\
        Gradient accumulation steps & 4 \\
        Block size for token-level & 1024 \\ 
        Block size for line-level & 924 \\
    \end{tabular}
    \caption{Model Hyperparameters.}
    \label{tab:1}
\end{table}

% During the training on \gls{mtl} models, we also have try different tasks weighing technique. The results are shown in session 5.3. 

% We improve the post-processing process from CodeXGLUE source code \footnote{https://github.com/microsoft/CodeXGLUE/tree/main/Code-Code/CodeCompletion-line}. As we observe that after finish line-predicting the existing CodeXGLUE method use function '.strip($\langle EOL \rangle$)' to truncate the $\langle EOL \rangle$ token. We found that this function doesn't remove only the given word, however every given characters from the beginning and the end of the original string \footnote{https://www.simplilearn.com/tutorials/python-tutorial/strip-in-python}. Therefore, it creates the false truncated results such as in Fig. \ref{}. 

% We propose to use '.replace($\langle EOL \rangle$,'')' for post-processing instead. Since in line-level prediction the $\langle EOL \rangle$ token is represented as the end, there will certainly have no $\langle EOL \rangle$ token in-between the results. Thus the replace function is more suitable for the truncation in this scenario. 

\subsection{Evaluation Measurement}
We evaluate our models based on the following evaluation measures: Accuracy (Acc) for token-level predictions; Exact Match (EM), Edit Similarity (ES), Mean Reciprocal Rank (MRR), \revise{R3.11}{BLEU, METEOR, and ROUGE} for line-level predictions.

% \our~performs two level of predictions: token-level prediction and line-level prediction. We evaluate token-level prediction with one metric: accuracy (Acc); and line-level with three metrics: exact match (EM), edit similarity (ES) and mean reciprocal rank (MRR).

% \subsubsection{Token-level prediction} %\textbf{Token-level}

\textbf{Accuracy (Acc)} is the proportion of correctness between predicted code tokens to the ground-truth tokens. 
% We use the proportion of correctness between predicted code tokens to the ground-truths as the evaluation for single token prediction. 

% \subsubsection{Line-level prediction}

\textbf{Exact Match (EM)} is similar to  Accuracy, but is evaluated at the line level, meaning that the whole predicted lines must be exactly matched with the ground-truth lines. 

% We call this measurement as exact match to be consistent with CodeXGLUE \cite{lu2021codexglue}.
% Similar to token-level prediction, we also measure the correctness (i.e., accuracy) of line-level prediction by comparing the whole predicted code line. We call this measurement as exact match to be consistent with CodeXGLUE \cite{lu2021codexglue} evaluation.

\textbf{Edit Similarity (ES)} uses a  Levenshtein distance~\cite{1966SPhD...10..707L} to measure the edit distance between the predicted lines and ground-truth lines. 
The Levenshtein distance is the minimum number of edits in characters (either an insertion, a deletion, or a replacement of a character) between the predicted line and the ground-truth line. 
% We use Levenshtein distance~\cite{1966SPhD...10..707L} to measure the edit similarity between predictions and ground-truths. The Levenshtein distance is the minimum number of edit in characters from one text to the other. The edit is defined by either an insertion, a deletion, or a replacement of a character. 

\textbf{Mean Reciprocal Rank (MRR)} evaluates the top-$R$ possible results using the multiplicative inverse of the rank of the first correct prediction. 
Formally, MRR is defined as:
% We evaluate the list of top-R possible results using multiplicative inverse of the rank of the first correct prediction. The rank is defined as:
\begin{equation}
    \label{eq:mrr}
    MRR=\frac{1}{|Q|}\sum_{i=1}^{|Q|}\frac{1}{rank_i}
\end{equation}
% \[ MRR=\frac{1}{|Q|}\sum_{i=1}^{|Q|}\frac{1}{rank_i} \]
\noindent, where $Q$ is the number of samples, and $rank_i$ is the rank of the correct prediction given by the model.
If the correct prediction exceeds rank $R$, then the reciprocal rank is 0.
In this paper, we use $R=5$.

% machine-translated text
\revise{R3.11}{\textbf{BLEU} evaluates how similar the predicted lines to the ground-truth lines using n-gram~\cite{papineni2002bleu}. 
In this paper, we use the cumulative 4-gram BLEU score (i.e., BLEU-4) from NLTK library\footnote{https://www.nltk.org/api/nltk.translate.bleu\_score.html}, with the weight of 0.25 for each of the 1-gram, 2-gram, 3-gram, and 4-gram score.}

% machine-translated text
\fix{\textbf{METEOR} evaluates how similar the the predicted lines to the ground-truth lines based on the harmonic mean of unigram precision and recall~\cite{banerjee2005meteor}.}

% machine-translated and summarization text
\fix{\textbf{ROUGE} measures the quality of the summary by counting the number of overlapping units such as n-gram, token sequences and token pairs between the predicted lines and the ground-truth lines~\cite{lin2004rouge}.
In this paper, we use ROUGE-L, which is the Longest Common Subsequence (LCS)~\cite{lin2004automatic} based statistics.}

\subsection{Baselines}

There exist various non-AST-based code completion approaches in CodeXGLUE~\cite{lu2021codexglue,radford2019language} and AST-based code completion approaches~\cite{kim2021code,izadi2022codefill,li2017code} in the literature.
To ensure that our evaluation is reasonably comprehensive, we consider a total of \revise{R3.15}{eight (8)} baselines with respect to two evaluation settings: (1) externally evaluate the prediction results through the CodeXGLUE leaderboard,\footnote{https://microsoft.github.io/CodeXGLUE/} and (2) internally evaluate the prediction results within our own setting.

For the CodeXGLUE evaluation setting, we compare our approach with CodeGPT-adapt, CodeGPT, GPT-2, Transformer (12L), and LSTM+BPE.
To do so, we apply our \our~to the testing set provided by CodeXGLUE for both token-level and line-level predictions.
Then, the prediction results are submitted to the CodeXGLUE team to obtain the results based on their evaluation setting.
\revise{R3.15}{Additionally, we also include the results from UniXcoder~\cite{guo2022unixcoder} for a comprehensive comparison, as the authors experimented on the same CodeXGLUE benchmark setting.}

For our own evaluation setting, we consider two AST-based approaches (i.e., Pointer Mixture Network~\cite{li2017code} and TravTrans~\cite{kim2021code}); two non AST-based approaches (i.e., GPT-2 and CodeGPT)\revise{R3.15}{; and a multi-modal pretrain model (i.e., UniXcoder)}.
We do not consider CodeFill~\cite{izadi2022codefill}, since the available replication package is not executable. 
We also do not consider Codex (i.e., a descendant of GPT-3 for source code) in our experiment due to the different levels of model parameter size. 
GPT-3, a base model of Codex, has 175B model parameters, which is 100x larger than the size of our GPT-2 based model which has only 117M model parameters.
Below, we describe the details of each approach.

% We also consider CodeFill model\footnote{https://github.com/saltudelft/codefill}
% as a baseline; however, the replication package of CodeFill is not available at the time of the experiments\footnote{We try sending an email to ask the authors, but there is still no updates on the replication package.}.

% \kla{we can even say that we compare with both AST-based and non AST-based approaches.}
% Apart from sending our model results to evaluate in CodeXGLUE benchmark~\cite{lu2021codexglue}, we compare our model with four state-of-the-art approaches to provide the concise evaluation.
% The baselines include both AST-based models: Pointer Mixture Network and TravTrans; and non AST-based models: GPT-2 and CodeGPT. 

% We run all baselines using their replication packages provided by authors. 

% The string and numerical literals in PY150 AST dataset of Pointer Mixture Network and TravTrans are masked with the similar pre-processing method to ours before the training process. 
% We also consider CodeFill model\footnote{https://github.com/saltudelft/codefill}
% as a baseline; however, the replication package of CodeFill is not available at the time of the experiments\footnote{We try sending an email to ask the authors, but there is still no updates on the replication package.}.
% \kla{codefill [AST], but the replication not available}

\begin{table*}[t!]
\centering
\caption{(RQ1) The results that appear in the CodeXGLUE leaderboard 
% as of 15 October 2022 filtered only Python
(\url{https://microsoft.github.io/CodeXGLUE/}).
% The evaluation results are calculated based on their ground-truth dataset.
% \kla{add ranking column? team name?}
% Compare results to the baseline in CodeXGLUE Benchmark. \kla{will check again after the table is revised. PyCoder to the top, sort from max to min. Double-check capitalization carefully, Make it neat, fits the table nicely to the column. add batch size column too. }
}
% \resizebox{\columnwidth}{!}{
\begin{tabular}{c|l|l|l|c|c|c}
    & & & & \multicolumn{2}{c}{\textbf{Line-level}} & 
    \multicolumn{1}{|c}{\textbf{Token-level}} \\
    \hline
    \textbf{Rank} & \textbf{Model} & \textbf{Team name} & \textbf{Date} & \textbf{EM} & \textbf{ES} & \textbf{Acc} \\
    \hline
    1 & \our-Hard & Monash University & 2022-10-13 & \textbf{43.91} & 71.74 & \textbf{76.93}\\
    \hline
    2 & \fix{UniXcoder} & Guo~\ea & 2022-03-08 (publish data) & 43.12 & \textbf{72.00} & - \\
    % \hline
    3 & CodeGPT-adapt & CodeXGLUE Team & 2020-08-30 & 42.37 & 71.59 & 76.60 \\
    4 & CodeGPT & CodeXGLUE Team & 2020-08-30 & 42.18 & 71.23 & 76.58 \\
    5 & GPT-2 & CodeXGLUE Team & 2020-08-30 & 41.73 & 70.60 & 75.90 \\
    6 & Transformer (12L) & CodeXGLUE Team & 2020-08-30 & 38.51 & 69.01 & 74.48 \\
    7 & LSTM + BPE & CodeXGLUE Team & 2020-08-30 & 23.77 & 56.26 & 61.94 \\
    % \hline
    % \our-Hard& \textbf{76.93} & \textbf{43.91} & \textbf{71.74} \\
\end{tabular}
% }
\label{tab:rq1 codexglue}
\end{table*}

% The evaluation results based on the CodeXGLUE leaderboard show that \our~outperform other baselines by \kla{X\%-Y\%}.

\begin{table*}[t]
\centering
\caption{(RQ1) The results of \our~when compared to existing approaches through our internal evaluation.
}
% \resizebox{\columnwidth}{!}{
\begin{tabular}{l|c|c|c|c|c|c|c}
    & \multicolumn{6}{c}{\textbf{Line-level}} & \multicolumn{1}{|c}{\textbf{Token-level}} \\
    \hline
    \textbf{Model} & \textbf{EM} & \textbf{ES} & \textbf{MRR} & \fix{\textbf{BLEU-4}} & \fix{\textbf{METEOR}} & \fix{\textbf{ROUGE-L}} & \textbf{Acc} \\
    \hline
    \our-Hard & \textbf{43.37} & \textbf{73.20} & \textbf{48.82} & \textbf{46.03} & \textbf{40.42} & \textbf{59.97} & \textbf{77.12} \\
    \hline
    \fix{UniXcoder} & 40.68 & 71.99 & 45.85 & 43.31 & 39.26 & 58.36 & - \\
    CodeGPT & 40.03 & 70.61 & 46.64 & 42.12 & 38.53 & 56.96 & 75.69 \\
    TravTrans & - & - & - & - & - & - & 75.50\\
    GPT-2 & 37.64 & 68.44 & 43.85 & 39.23 & 36.89 & 54.32 & 73.89 \\
    PMN & - & - & - & - & - & - & 69.02\\
\end{tabular}
% }
\label{tab:rq1 our}
\end{table*} 

\begin{itemize}
    \item \textbf{Pointer Mixture Network (PMN)}, proposed by Li~\ea~\cite{li2017code}, is an LSTM-based code completion
    % \kla{code completion--to make it context-specific, not generic LSTM model } model 
    % ?????\kla{fix the rest, same pattern as below} 
    leveraging AST information for syntactic structures.
    The model is designed with pointer networks to mitigate the OOV problems in code completion.
    % The model learns to generate the next token from either within vocabulary word or regenerate an Out-of-Vocabulary word from local context through the pointer component.
    Their replication package is available on Github\footnote{https://github.com/jack57lee/neuralCodeCompletion} and also in Pytorch version\footnote{https://github.com/oleges1/code-completion}.

    \item \textbf{TravTrans}, proposed by Kim~\ea~\cite{kim2021code}, is a transformer-based model that considers the syntactical structure of source code via AST information.
    Their replication package is available on GitHub\footnote{https://github.com/facebookresearch/code-prediction-transformer}.
    
    \item \textbf{GPT-2}, proposed by Radford~\ea~\cite{radford2019language}, is a GPT-2-based model for text generation tasks. 
    The GPT-2 model is first pre-trained on millions of English web pages (the WebText corpus) to build a language model through self-supervision learning without any explicit labels. 
    The model is available on HuggingFace\footnote{https://huggingface.co/gpt2}.
    
    % The tokenizer use in this work is BPE~\cite{sennrich2015neural}.
    \item \textbf{CodeGPT}, proposed by Lu~\ea~\cite{lu2021codexglue}, is a GPT-2-based model for source code generation.
    The CodeGPT model is a GPT-2 model that is pre-trained on a monolingual python source code from CodeSearchNet~\cite{husain2019codesearchnet} dataset.
    % \kla{CSN has many PLs. are you sure only python?}.
    The model is available on HuggingFace\footnote{https://huggingface.co/microsoft/CodeGPT-small-py}. 

    \revise{R3.15}{
    \item \textbf{UniXcoder}, proposed by Guo~\ea~\cite{guo2022unixcoder}, is a Transformers-based unified cross-modal pre-trained model for source code generation. 
    The UniXcoder model is pretrained on three types of language model tasks: masked language model, unidirectional language model, and denoising objective, and two types of inputs: comment and flattened AST.
    The pretrain datasets are six programming languages from CodeSearchNet dataset~\cite{husain2019codesearchnet}, and natural text from C4 dataset~\cite{raffel2020exploring}.
    Their replication package is available on Github\footnote{https://github.com/microsoft/CodeBERT/tree/master/UniXcoder}.
    }
    
    % In CodeXGLUE benchmark dataset from Lu~\ea~\cite{lu2021codexglue} proposed baseline: CodeGPT for code generation tasks.
    % CodeGPT has the similar transformer-based model as GPT-2; however, the model is pre-trained on monolingual python source code dataset i.e. CodeSearchNet~\cite{husain2019codesearchnet}.
    % The BPE tokenizer was also newly train on source code data.
    % The model is available on HuggingFace\footnote{https://huggingface.co/microsoft/CodeGPT-small-py}
\end{itemize}

\section{Experimental Results}\label{sec:results}

In this section, we present the experimental results according to our four research questions (RQs).

\begin{table*}[t]
    \centering
    \caption{(RQ2) The results of \our~when using various multi-task training strategies. For a fair comparison with other multi-task training strategies, we do not put any weights between tasks on \our-Hard (i.e., \emph{No Weight}).
    % \kla{I feel type acc should be dropped. irrelevant, and never talked about it before.}
    }
    % \resizebox{\columnwidth}{!}{
    \begin{tabular}{l|c|c|c|c|c|c|c}
        & \multicolumn{6}{c}{\textbf{Line-level}} & 
        \multicolumn{1}{|c}{\textbf{Token-level}}\\
        \hline
        \textbf{Model} & \textbf{EM} & \textbf{ES} & \textbf{MRR} & \fix{\textbf{BLEU-4}} & \fix{\textbf{METEOR}} & \fix{\textbf{ROUGE-L}} & \textbf{Acc} \\%& Type Acc \\
        \hline
        \our-Hard & \textbf{42.83} & \textbf{72.82} & 48.42 & \textbf{45.52} & \textbf{40.05} & \textbf{59.40} & \textbf{77.00} \\ %& \textbf{82.49} \\
        \our-IFN & 42.04 & 71.92 & \textbf{48.54} & 44.18 & 39.40 & 58.41 & 76.52 \\ %& 0.0 \\
        \our-Soft & 38.29 & 69.11 & 44.66 & 39.85 & 37.36 & 55.24 & 74.77 \\ %& 80.68 \\
    \end{tabular}
    % }
    \label{tab:rq2}
\end{table*}

\begin{table*}[t]
    \centering
    \caption{(RQ3) The results of different task weighing parameters for Hard Parameter Sharing only as \our-Hard performs best.}
    % \resizebox{\columnwidth}{!}{
    \begin{tabular}{l|c|c|c|c|c|c|c}
        \textbf{Task's Weight} & \multicolumn{6}{c}{\textbf{Line-level}} & 
        \multicolumn{1}{|c}{\textbf{Token-level}}\\
        \hline
        \textbf{Type:Code} & \textbf{EM} & \textbf{ES} & \textbf{MRR} & \fix{\textbf{BLEU-4}} & \fix{\textbf{METEOR}} & \fix{\textbf{ROUGE-L}} & \textbf{Acc} \\ %& Type Acc \\
        \hline
        No Weight & 42.83 & 72.82 & 48.42 & 45.52 & 40.05 & 59.40 & 77.00 \\% & 82.49 \\
        1 : 9 & \textbf{43.37} & \textbf{73.20} & 48.82 & 46.03 & \textbf{40.42} & \textbf{59.97} & \textbf{77.12} \\ %& 81.16 \\
        2 : 8 & 43.08 & 73.01 & 48.73 & \textbf{46.12} & 40.37 & 59.69 & \textbf{77.12} \\ %& 81.72 \\
        3 : 7 & 42.95 & 72.94 & 48.56 & 45.80 & 40.27 & 59.57 & 77.10 \\% & 82.05 \\
        4 : 6 & 42.84 & 73.03 & 48.55 & 45.75 & 40.31 & 59.68 & 77.05 \\% & 82.29 \\
        5 : 5 & 42.94 & 72.69 & \textbf{49.53} & 45.39 & 39.99 & 59.42 & 76.99 \\% & 82.47 \\
        6 : 4 & 42.37 & 72.29 & 49.10 & 44.77 & 39.78 & 59.09 & 76.88 \\% & 82.65 \\
        7 : 3 & 42.28 & 72.27 & 47.82 & 44.83 & 39.68 & 58.89 & 76.70 \\% & 82.77 \\
        8 : 2 & 41.19 & 71.68 & 46.76 & 43.79 & 39.25 & 58.07 & 76.23 \\% & 82.75 \\
        9 : 1 & 39.77 & 70.52 & 46.34 & 41.80 & 38.34 & 56.66 & 75.45 \\% & \textbf{82.83}
    \end{tabular}
    % }
    \label{tab:rq3 hard-share}
\end{table*}

\subsection*{\textbf{(RQ1) \rqone}}

% \kla{Wannita, please compute the \% in the blank throughout the paper. Thanks.}

% Compare results to the state-of-the-art baselines. \kla{will check again after the table is revised. PyCoder to the top, sort from max to min. Double-check capitalization carefully, Make it neat, fits the table nicely to the column. add batch size column too. }

\indent \textbf{\our.} Among our comprehensive investigation, the best setting for \our~is to train with the hard parameter sharing strategy (\our-Hard), a task weight of 9:1 (code:type) using a Beam Search as a decoding method.
We use this setting as a reference for comparison with other approaches throughout the paper.

% \noindent \textbf{Results.} 
\textbf{\our~achieves the first rank on the CodeXGLUE leaderboard for the code completion task} (as of 13 October 2022, see Table~\ref{tab:rq1 codexglue}).
% \kla{ in the website, it shows line first, then token. for our case, there is a bigger improvement for line, than token. why don't we sell line first, then token? thought?}
We find that \our~achieves an accuracy of 76.93\% for the token-level predictions, while achieving an exact match of 43.91\% for the line-level predictions.
The evaluation results confirm that  \our~is more accurate than other baselines by 0.43\%-24.25\% for token-level predictions and 3.63\%-84.73\% for line-level predictions.
% \kla{may talk about why small improvement - could be due to batch size, limited GPU capacity, etc... at least, still outperform and}

Similarly, \textbf{\our~outperforms existing AST-based and non-AST-based code completion approaches}, according to our own setting.
Table~\ref{tab:rq1 our} shows that \our~achieves an accuracy of 77.12\% for the token-level predictions, while achieving an exact match of 43.37\% for the line-level predictions.
For the token-level predictions (Acc), we find that \our~is more accurate than Pointer Mixture Network by 11.74\%, GPT-2 by 4.37\%, TravTrans by 2.15\%, and CodeGPT by 1.89\%. 
This finding indicates that \our~that is syntax-aware and on-the-fly performs better than a code completion approach that is either syntax-aware alone or on-the-fly alone.
It is worth noting that the accuracy of \our-Hard, CodeGPT, and GPT-2 achieved for the CodeXGLUE leaderboard is slightly different from the accuracy of those that are run in our experimental setup.
The difference that we observed has to do with the dataset used in CodeXGLUE and our experiment.
In CodeXGLUE, they removed the indentation, while the dataset used in our experiment preserved the original indentation.
To mimic the practical deployment scenario, we opt to preserve the original indentation.

% \kla{will talk about the small diff value btw codexglue and our own here why}

In addition, \textbf{the token-type information can improve the line-level code completion task by %8.34\%
\fix{6.61\%}-15.22\%.}
For the line-level predictions (EM), we find that \our~is more accurate than GPT-2 by 15.22\%, CodeGPT by 8.34\%, and \revise{R3.15}{UniXcoder by 6.61\%}.
This finding indicates that the use of token-type information that is largely ignored by the literature can also improve line-level predictions by 
%8.34\% 
\fix{6.61\%} to 15.22\%, confirming that the token-type information is useful to improve the performance of line-level code completions.

Finally, when comparing \our~with the existing AST-based code completions (i.e., TravTrans and Pointer Mixture Network), we find that the existing AST-based code completions are designed for the token-level predictions only.
Thus, the line-level predictions cannot be performed, highlighting the limitations of the AST-based approaches that require AST information at the inference time, while demonstrating the benefits of our approach that consider the token-type information (i.e., \emph{syntax-aware}), while can still predict code at any points of time (i.e., \emph{on-the-fly}).

\begin{tcolorbox}

% To answer this RQ, we build SynComp from the best combination of experiments on model technique (RQ2), task weighing parameters (RQ3) and decoding methods (RQ4) i.e. hard parameters sharing model with weighing code to type tasks as 9:1, and beam search as decoding methods.See the details of each experiment result in the following RQs.
\emph{\textbf{RQ1 Summary.} 
\fdone
% \our~surpasses all the state-of-the-art code completion models.
% The line-level exact match is improved by 8.34\%-15.22\%, while the token-level accuracy is improved by 1.89\%-11.74\%.
% \our~also receives the first place in CodeXGLUE’s python code completion benchmark. The results indicate that the token type syntactic information can be beneficial as a supporting data.
}
\end{tcolorbox}

% \subsection{RQ2: What is the impact of the training strategies on the performance of our approach}

% To answer this research question, we compare the best results of each model techniques after tuning weighing (see session 5.3). The results are shown in Table \ref{tab:rq2}. From the table, hard parameters sharing with manual weight of 9:1 perform the best in both token-level and line-level prediction. This is also the method we decide to use sending results to CodeXGLUE competition. Second place is \gls{ifn} method. Even though \gls{ifn} method give comparable results to hard parameters sharing model. When we train further to the optima, the \gls{ifn} model are struggled on forgetting catastrophe \cite{} which cause the model to perform worse than hard parameters sharing later.   

% if no-weight results -> write about even if IFN is slightly better but when training further the results drop because of the forget catastrophe

% \textbf{Approach.} To answer this research question, we compare the results of each model training techniques: hard parameters sharing (MTL), soft parameters sharing (MTL), and intermediate fine-tuning (STILTs); before tuning the task weighing parameters. 

\subsection*{\textbf{(RQ2) \rqtwo}}

Table~\ref{tab:rq2} presents the results of \our~when using various multi-task training strategies.

\textbf{Hard parameter sharing (\our-Hard) as a multi-task training strategy performs the best.}
Table~\ref{tab:rq2} shows that different multi-task training strategies have an impact on the performance of \our~for both token-level and line-level predictions.
Particularly, we observe that \our~with hard parameter sharing achieves an exact match of 42.83\%, while \our~with software parameter sharing achieves an exact match of 38.29\%. 
The 4.54\% difference (i.e., max-min) confirms the impact that the training strategies have on the performance of \our.
In addition, our results are contradictory to Izadi~\ea~\cite{izadi2022codefill} who found that soft parameter sharing performs best for code completion.
% , while being in line with Liu~\ea~\cite{liu2022unified} who found that hard parameters sharing performs best.
This finding highlights the importance of investigating various choices of multi-task training strategies for code completions, instead of following prior suggestions or practices.

Different from Izadi~\ea~\cite{izadi2022codefill}, our \our-Hard is designed to take both sequences of code tokens and their types as inputs one-by-one at a time and simultaneously learn with the same loss functions that are optimized together within the same model.
With this method, the inputs can be detached from each other at the inference phase, resulting in better performance confirmed by our results.
Nonetheless, the high-performing hard parameter-sharing training strategy (\our-Hard) has to do with the benefits of the tight relationship between the learning tasks (i.e., code predictions and type predictions).
Since token types are directly aligned with the same sequence of code tokens, these two pieces of information have a tight relationship.
Therefore, \our-Hard, which completely shares the model’s weights and parameters between tasks, gains the most benefit from the shared relationship between the code and type information.
However, the soft parameter sharing model (\our-Soft) learns each task separately, making the learning process between two related tasks harder, resulting in sub-optimal performance.

\begin{tcolorbox}
\emph{\textbf{RQ2 Summary.}} 
\fdtwo

% highlights the importance of investigating various choices of multi-task training strategies for code completions, instead of following prior suggestions or practices

% The best multi-task training technique for our setting is \our-
% IFN: Intermediate Fine-tuning (\our-IFN) and MTL: Soft Parameter Sharing (\our-Soft).
% Particularly, the exact match in line-level prediction is vary by 4.54\%.

\end{tcolorbox}

\subsection*{\textbf{(RQ3) \rqthree}}
% \subsection{RQ3: What is the impact of the task weighting parameters in multi-task learning on the performance of our approach?}
% \begin{table}[]
%     \centering
%     \begin{tabular}{l|c|c|c|c|c}
%         Model & Code Acc & Type Acc & \gls{em} & \gls{es} & \gls{mrr} \\
%         \hline
%         Random & & & & &\\
%         Random with control & & & & &\\
%         Manual 7:2:1 & & & & &\\
%     \end{tabular}
%     \caption{The sensitivity of weighing for soft parameters sharing model performances.}
%     \label{tab:rq3 soft-share}
% \end{table}

\begin{table*}[t]
    \centering
    \caption{(RQ4)  The results of different decoding methods for Hard Parameter Sharing (1:9). For any sampling methods, we report both the Mean and its standard deviation (SD).}
    % \resizebox{\columnwidth}{!}{
    \begin{tabular}{l|l|c|c|c|c|c}
        & & \multicolumn{5}{c}{\textbf{Line-level}} \\
        \hline
        \textbf{Library} & \textbf{Method} & \textbf{EM} & \textbf{ES} & \fix{\textbf{BLEU-4}} & \fix{\textbf{METEOR}} & \fix{\textbf{ROUGE-L}} \\
        \hline
        \multirow{2}{*}{CodeXGLUE}
        & Beam Search ($b$=5) & \textbf{43.37} & 73.20 & 46.03 & 40.42 & 59.97 \\ 
        & Greedy & 41.47 & \textbf{73.95} & \textbf{47.83} & \textbf{42.34} & \textbf{60.46} \\ 
        \hline
        \multirow{6}{*}{HuggingFace}
        & Beam Search ($b$=5) & \textbf{41.52} & 72.85 & 46.8 & 42.04 & 59.62 \\
        & Greedy & 41.48 & 73.96 & 47.83 & \textbf{42.35}& 60.47 \\ 
        & Sampling & 33.80 (0.16) & 68.72 (0.11) & 40.80 (0.18) & 39.32 (0.08) & 54.18 (0.16) \\ 
        & Sampling with Temp ($temp$=0.1) & 41.38 (0.10) & \textbf{73.98 (0.03)} & \textbf{47.91 (0.07)} & 42.31 (0.04) & \textbf{60.48 (0.04)} \\ 
        & Top-K Sampling ($k$=3) & 35.34 (0.16) & 70.43 (0.11) & 43.01 (0.12) & 40.15 (0.09) & 56.43 (0.11) \\ 
        & Top-P Sampling ($p$=0.1) & 41.48 (0.01) & 73.95 (0.01) & 47.82 (0.03) & \textbf{42.35 (0.01)} & 60.46 (0.01) \\
    \end{tabular}
    % }
    \label{tab:rq4}
\end{table*}

% \textbf{Approach.} To answer this research question, we choose the hard parameters sharing model to experiment on the weighing sensitivity as it is the best model from RQ2. We manually tune the model with different weights according to equation~\ref{eq:rq3} where $\alpha$ is the weight.

% \textbf{Results.} 
Table~\ref{tab:rq3 hard-share} presents the results of different task weighing parameters for \our-Hard.

\textbf{\our~is generally robust to the task weighting parameters, achieving comparative (without task weighting) or better (with task weighting) performance when compared to the baselines.}
Table~\ref{tab:rq3 hard-share} shows that when varying the task weighting parameters (Type:Code) from 1:9 to 9:1, our \our~achieves an exact match between 41.19\% to 43.37\%, which is still greater than the existing approaches (i.e., 40.03\% for CodeGPT and 37.64\% for GPT-2) with an exception for the weighting of 9:1.
Although the task parameters are not weighted (cf. \emph{No Weight}), our \our~still achieves an exact match of 42.83\%, which also outperforms the existing approaches.
In line with the other measures for both line-level and token-level predictions, this finding confirms that by adding token-type information by at least a small weighting of 10\%, our \our~often performs better than the existing approaches.
% \kla{need to mention the gradient thing}
This means that the task objectives of \our~rarely suffered from conflicting gradients (i.e., the gradients of different task objectives are not aligned leading to the sub-optimal performance in the average gradient) showing that type prediction and code prediction are correspondent and beneficial to each other.
In our setting, the best task's weight is 1:9 for the type prediction task to the code prediction task.

\begin{tcolorbox}
\emph{\textbf{RQ3 Summary.}
\fdthree{}
% The results show that \our~is robust to the task weighing parameters; additionally, the model can perform slightly better in both level predictions when tuning the weights.
% The best task’s weight for our work is 9:1 for code to type prediction task.
}
\end{tcolorbox}

\subsection*{\textbf{(RQ4) \rqfour}}
% \subsection{RQ4: What is the impact of the decoding methods on the performance of our approach?}

% To recommend code review comments, Com- mentFinder leverages the cosine distance to retrieve the ten nearest changed methods and GPM to identify the �� most similar changed methods. Although other distance and text similarity metrics are available, they may provide different results. Yet, little is known about the impact that these simi- larity techniques can have on the effectiveness and efficiency of our CommentFinder. Hence, we formulate this RQ to further investigate the possible impact when the distance metrics or text similarity metrics in CommentFinder are changed.

\begin{figure*}
    \centering
    \includegraphics[width=\textwidth]{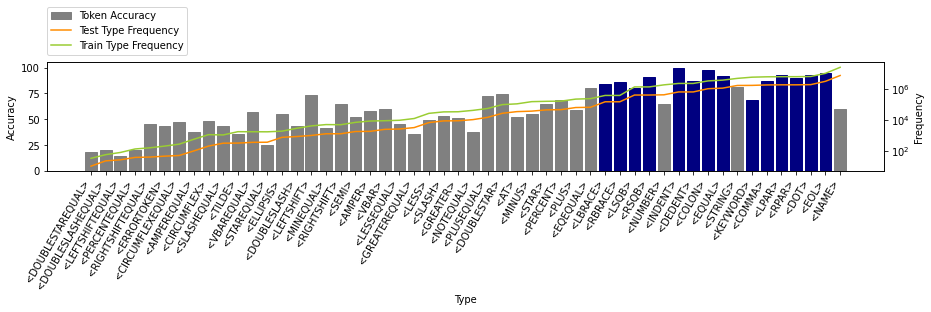}
    \caption{The chart of the token-level code prediction accuracy in token type granularity sorted by the type frequency from small (left) to large (right). The tokens related to syntax types is represented in blue color.}
    \label{fig:type_barplot}
\end{figure*}

% \textbf{Approach.} To answer this research question, we study the difference between following decoding methods. 

% \input{sections/decodingMethods}
% \textbf{Results.}
% We also compare Beam search from two libraries: CodeXGLUE and HuggingFace.
% For Greedy method, we use Beam search with beam size $b$=1.
% Note that the model we use here is the hard parameter sharing model with task weighing parameter equal to 9:1 (code to type) since it is the best combination from experiments in RQ2 and RQ3.
Since decoding methods are specially designed for generating code predictions as a sequence (i.e., not an individual code token), the rest of this RQ will focus on the line-level predictions only, not the token-level predictions.
% is to handle a search space for a sequence prediction,
% i.e. line-level prediction.
% \kla{???, you can fill}.
% Thus, we will not focus on the impact of decoding methods on the token-level predictions.
% We perform the experiments for this RQ in the line-level prediction since selecting only the highest probability token is suitable for predicting a token\kla{this is the explanation for token-level, not why choosing line-level}, however, it might be a sub-optimal for a sequence.
We note that some decoding methods (i.e., Beam Search and Sampling with a probability shaping function) require parameter settings to be specified.
Thus, we experiment with the following parameters: a beam size ($b$) of $\{3, 5, 10, 16, 50\}$ for Beam Search, a temperature ($temp$) of $\{0.05, 0.1, 0.3, 0.5, 0.7, 0.9\}$ for Sampling with Temperature, a top-k ($k$) of $\{3, 5, 10, 50, 100\}$ for Top-K Sampling, and a top-p ($p$) of $\{0.05,0.1,0.3,0.5,0.7,0.9\}$ for Top-P Sampling.
For the Sampling approaches, we repeat the experiment five (5) times with different seed numbers to ensure the robustness of the results.
Thus, we present the results using the average of the distribution and its standard deviation (SD).
% ; then, the final reported number are the average score and the .
% ......
% \kla{why line level?} 
% We adjust different sizes of  beam size ($b$), top-k ($k$), top-p ($p$) and temperature ($temp$) \kla{why these params?} to our best effort using $b \in \{3, 5, 10, 16, 50\}$, $temp \in \{0.05, 0.1, 0.3, 0.5, 0.7, 0.9\}$,  $k \in \{3, 5, 10, 50, 100\}$, and $p \in \{0.05,0.1,0.3,0.5,0.7,0.9\}$; then, select the best value of each method to compare in this RQ (i.e. $b$=5, $k$=3, $p$=0.1 and $temp$=0.1).
Since Beam Search and Greedy methods are available in both CodeXGLUE and HuggingFace libraries with different implementations, we also evaluate decoding methods using both libraries.
Finally, we experiment with a total of 102 variants of 6 decoding methods, i.e., $\bigl($2*libraries $\times$ (1*Greedy + 5*BeamSearch)$\bigr)$ + $\bigl($5 repeats $\times$ (1*Sampling, 6*Temp, 5*$k$, 6*$p$)$\bigr)$. 

% Thus, we also compare the results to see the impact of decoding methods from the different libraries. 
% We also compare the results between two different libraries: CodeXGLUE and HuggingFace. \kla{why 2 libs?}
% different decoding methods has impact, different library has impact, best is beam search > sampling 

\textbf{Beam Search performs the best, while Sampling performs the worst.}
Table~\ref{tab:rq4} shows that there is a great performance difference of \our~when different decoding methods are used.
For example, Beam Search(CodeXGLUE) generally achieves an exact match of 43.37\%, while Sampling achieves an exact match of 33.80\%, confirming that the decoding methods have a substantial impact on the performance of \our for line-level code completion.
In addition, we find that not only the methods but different libraries with different implementations also produce different results.
In particular, when comparing Beam Search between CodeXGLUE and HuggingFace libraries (see Table~\ref{tab:rq4}), we find that Beam Search from the CodeXGLUE library achieves an exact match of 43.37\% (used by \our), which is greater than that from the HuggingFace library.
This finding suggests that future studies should use Beam Search(CodeXGLUE) for code completion and should report the library used for decoding methods for better reproducibility and replicability details.

% Moreover, the different decoding method implexact matchentations also impact the results.  
% Comparing between two Beam Search implementations (i.e., HuggingFace and CodeXGLUE), the exact match is vary from 41.52\% to 43.37\%.
% \kla{methods / libs - 2 points are mixing - hard to read}
% Table.~\ref{tab:rq4} indicates that considering only HuggingFace library results, there are a significant relative difference around 22.84\% %7.72\% 
% (from 33.80\% to 41.52\%) on the Exact Match and 7.64\% %5.26\% 
% (from 68.72\% to 73.98\%) on the edit similarity between Sampling method and Beam search method respectively.

% \textbf{Our best suggestion is to use Beam search or Greedy search as they are ones of the best performance which also do not require a random seed sampling.}

We find that Sampling is the lowest-performing decoding method, while advanced Sampling (i.e., Sampling with Probability Shaping) tends to perform better, depending on the specified parameter settings.
Through the comprehensive investigation, Top-P sampling performs best when $p$=0.1, and Sampling with Temp performs best when $temp$=0.1.
These optimal parameter settings are domain and context-specific to code completion, which are different from Holtzman~\ea~\cite{holtzman2019curious} who recommend $temp\in[0.5,1]$, $k\in[1,100]$, $p\in[0.9,1)$ for the text generation tasks.
The optimal setting that we achieved for code completion that is different from the recommendations in the NLP text generation field suggests that researchers should experiment with various parameter settings for the problem that tackle, instead of solely relying on suggestions or recommendations from prior work.

\begin{tcolorbox}
\emph{\textbf{RQ4 Summary.} 
\fdfour
% We found that the different decoding methods and libraries have impacts on the quality of predictions, and Sampling methods may not be suitable for code completion task. The best decoding method in our setting is Beam search with beam size $b$=5.
}
\end{tcolorbox}

\begin{figure}
    \centering
    \includegraphics[width=\columnwidth]{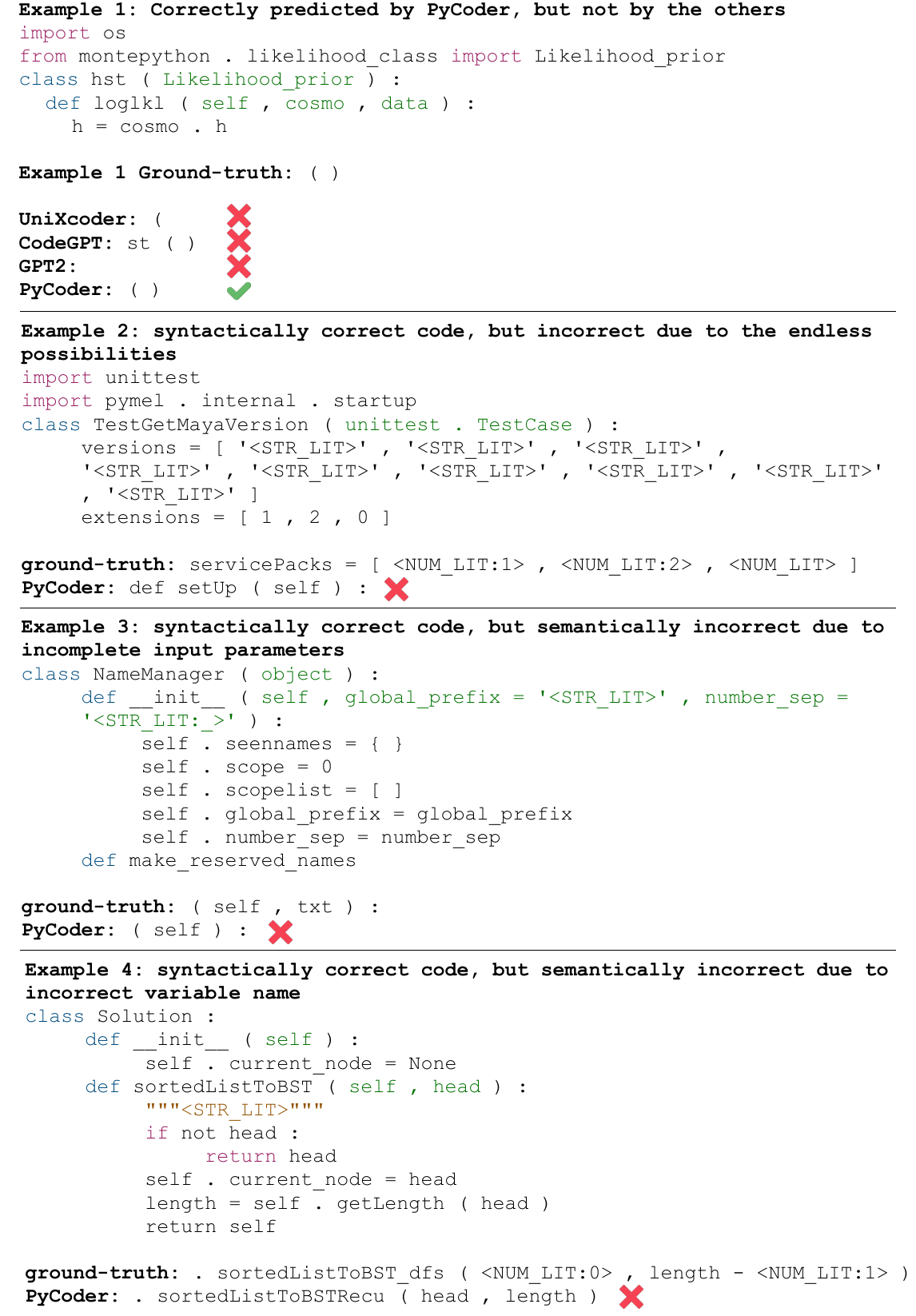}
    \caption{Examples of line-level predictions: (1) Correctly predicted by PyCoder, but not by the others; (2) Syntactically correct, but still incorrect due to endless possibilities; and (3,4) syntactically correct, but semantically incorrect.}
    \label{fig:samples}
\end{figure}

\revise{R2.9}{
\section{Discussion}
\label{sec:discussion}

\subsection{How many line-level predictions that our PyCoder can correctly predict while others cannot?}

To answer this question, we perform additional analysis on the line-level predictions between PyCoder and the state-of-the-art approaches (i.e., CodeGPT and GPT2 that PyCoder built upon).
We use a Venn diagram to visualize the number of correct (i.e., exact match) and incorrect predictions at the line level for each of the three approaches (see Figure~\ref{fig:venn3}). 

\textbf{We find that 11.5\% ($\frac{499}{4,337}$) of the line-level predictions can be correctly predicted by PyCoder while others cannot.}
Figure~\ref{fig:venn3} shows that, among the 10,000 line-level samples in the testing set, there are 4,878 samples that can be correctly predicted by one of the approaches, meaning that 5,122 samples cannot be correctly predicted by any of these three approaches.
Among the correct predictions, PyCoder can correctly predict the majority of the samples (i.e., 4,337), accounting for 88.9\% of the total correct predictions.
Most importantly, among these, 11.5\% ($\frac{499}{4,337}$) of samples can be accurately predicted by PyCoder while others cannot, highlighting the various key strengths of our approach that others do not have.

% \textbf{Venn diagram.} 
% In Figure~\ref{fig:venn3}, we display the relationship of exact match results on line-level code completion from \our, CodeGPT, and GPT-2.
% While areas within the circles represent the number of exact match results to the ground truth, the overlapped curves depict the mutual exact match results with other models.
% The figure indicates that although most of the exact match results are mutually inclusive, \our~has the highest frequency of unique correctness (i.e., 499 samples from 4,878 correct predictions).

\begin{figure}[t]
    \centering
    \includegraphics[width=0.8\columnwidth]{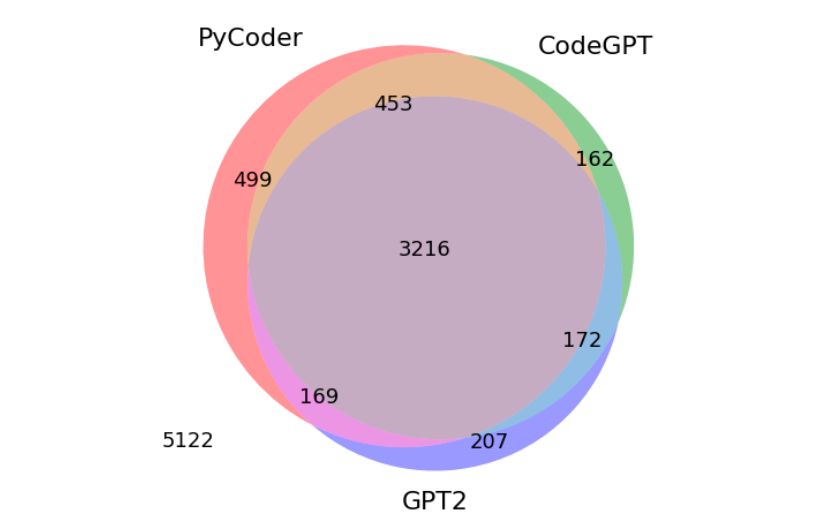}
    \caption{The Venn diagram of the exact match results on 10,000 samples of line-level code prediction from different models.}
    \label{fig:venn3}
\end{figure}

\subsection{Does PyCoder predict syntax-related tokens more accurately than the others?}

The key strength of PyCoder is based on the multi-task learning that combines token type information, while others (e.g., CodeGPT and GPT-2) don't.
Since \our~is specifically designed to incorporate token-type information, it is likely that PyCoder can predict syntax-related tokens more accurately than the others.
To answer this question, we perform additional analysis to investigate the relationship between the accuracy of code token predictions for each token type and the frequency of each token type that appears in the training and testing dataset (see Figure~\ref{fig:type_barplot}).

\textbf{We find that syntax-related types of tokens tend to be more accurate than other types of tokens (e.g., operational tokens, boolean and logical expressions, strings, and numbers).}
The difference in accuracy could be due to the amount of data in the training/testing dataset.
Figure~\ref{fig:type_barplot} shows that tokens related to syntax types (i.e., LPAR, RPAR, COLON, KEYWORD, INDENT, DEDENT, EOL) generally achieve an accuracy of  68.35\%-100.00\%, where these types account for 58.50\% and 58.33\% of the training and testing datasets, respectively.
% NUMBER 64.65
% ERRORTOKEN 43.75
On the other hand, operation-related tokens (e.g., PLUS, STAR, GREATER, NOTEQUAL) tend to be less accurate than syntax-related tokens, since these operation-related tokens tend to have less amount of tokens in the dataset.
The relationship between the code token accuracy and its frequency is also confirmed by Spearman's Rank Correlation of 0.85 (\emph{high}, $p$-value = 1.59$\times10^{-15}$), suggesting that more data in the training dataset may improve the code token predictions that are less frequent in the dataset.
This suggests that the performance of \our~may be dependent on the amount of dataset (could be either training or testing).

% its code token accuracy and the token type frequency
% In this section, we hypothesize that an accuracy performance might reflect on a different size of token type frequency. 
% \kla{why do we need to discuss?, let's start with what do we want to investigate, and why? need motivation} 
% Then, we present the token-level accuracy in another aspect showing a relationship between the code token accuracy and the token type frequency.
% To do so, we analyze the accuracy of code token predictions for each token type and the frequency of each type that appears in the training and testing dataset (see Figure~\ref{fig:type_barplot}).

Figure~\ref{fig:samples} presents an example line-level prediction that is correctly predicted by PyCoder, but not by the others.
Example 1 is a Python code snippet where the input line is an attribute call (\texttt{h = cosmo.h}), expecting to complete  \texttt{()} according to the ground truth.
When analyzing the output from the other approaches, we find that they may generate incomplete code tokens, causing syntax errors.
On the other hand, our PyCoder that learns with token types can accurately complete the attribute call with \texttt{()} while others cannot, demonstrating the effectiveness of PyCoder that considers token type information during multi-task learning.

\subsection{Why some predictions of our PyCoder are incorrect?}

To answer this question, we perform an error analysis to investigate the predictions of PyCoder that are incorrect when compared to the ground truth.
We first start from the group of incorrect predictions (i.e., 5,122 samples in Figure~\ref{fig:venn3}).
After a manual analysis of the random samples, we found the following two  patterns of incorrect predictions.

\textbf{PyCoder can generate syntactically correct code, but still incorrect due to the endless possibilities (e.g., after a new line).}
Example 2 demonstrates an example of syntactically correct generated code by PyCoder, but still incorrect due to the endless possibilities (e.g., after a new line).
For Example 2, the input is \texttt{extensions = [1,2,0]} with a new line, where the model is expected to complete the line after. 
As the model is expected to predict a new line, the possibility is endless. 
The ground truth is related to a variable declaration (\texttt{servicePacks = ...}), while PyCoder generates a new method name (\texttt{def setUp}).
While both recommendations are syntactically correct, but the recommended line from PyCoder still does not match the ground truth.
Therefore, it remains difficult for a model to complete the next line that is independent of the line before.

\textbf{PyCoder can generate syntactically correct code, but semantically incorrect (e.g., incomplete input parameter, incorrect method name).}
Examples 3 and 4 demonstrate an example of syntactically correct generated code by PyCoder, but semantically incorrect.
For Example 3, the input is \texttt{return self} where the model is expected to complete this line as \texttt{(self, txt):}.
However, PyCoder incorrectly completes this line as \texttt{(self):} due to the missing of an input parameter (\texttt{txt}).
For Example 4, the input is \texttt{def make\_reserved\_names} where the model is expected to complete this line as \texttt{.sortedListToBST\_dfs( <NUM\_LIT:0> , length-<NUM\_LIT:1>)}.
However, PyCoder incorrectly completes this line as \texttt{.sortedListToBSTRecu(head, length)} due to the incorrect method name (i.e., \texttt{sortedListToBSTRecu}).
}

% \begin{figure}
%     \centering
%     \includegraphics[width=\columnwidth]{figures/error_analysis.pdf}
%     \caption{Examples of \our~incorrect prediction for line-level code prediction on PY150 test set.}
%     \label{fig:error_anaysis}
% \end{figure}

% \begin{figure}
%     \centering
%     \includegraphics[width=0.5\columnwidth]{figures/venn2_codegpt.png}
%     \caption{The Venn diagram of exact match results on 10,000 samples of the line-level code prediction.}
%     \label{fig:venn2}
% \end{figure}

\fix{\revise{R3.3}{\section{Related Work}}
\label{sec:sec_relatedwork}

In this section, we mainly discuss related work in multi-task learning for code completion and highlight the novelty of PyCoder with respect to the existing work.

\begin{table*}[t]
    \centering
    \caption{The difference between PyCoder and the existing multi-task learning code completion.
    }
    \label{tab:mtl_compare}    
    \resizebox{\linewidth}{!}{
    \begin{tabular}{l|c|c|c|c|c|c|c}
        \hline
         & \multicolumn{2}{c|}{\textbf{\makecell{Training/Testing \\ Mechanism}}} 
         & \multicolumn{3}{c|}{ 
         \textbf{\makecell{Multi-task Learning \\ (Training Objectives)}}} 
         & \multicolumn{2}{c}{\textbf{\makecell{Sources \\ of Information}}} \\
        \hline
        \textbf{Approach} & \textbf{Fine-tuning} & \textbf{Inference} & \textbf{Hard share} & \textbf{Soft share} & \textbf{IFN} & \textbf{Semantic} & \textbf{Syntactic} \\
        \hline
        \our~(ours) & Code + Token Type & Code Only & \multicolumn{3}{c|}{\makecell{Two tasks: \\
        (1) Next Token Code prediction \\
        (2) Next Token Type prediction \\
        }} & Code & Standard Token Type \\ 
        \hline
        CugLM~\cite{liu2020multi}
        & Code Only & Code Only & \makecell{Three tasks: \\
        (1) Masked bidirectional LM \\
        (2) Next code segment prediction \\
        (3) Unidirectional LM
        } & N/A & N/A & Code & N/A \\ 
        \hline
        UMTLM~\cite{liu2020self, liu2022unified}
        & Code + AST & Code + AST & \makecell{Two tasks: \\
        (1) Next AST node type prediction \\
        (2) Next AST node value prediction} & N/A & N/A & Code & AST \\
        \hline
        CodeFill~\cite{izadi2022codefill}
        & Code + AST & Code + AST & \multicolumn{2}{c|}{\makecell{Three tasks: \\
        (1) Next AST token-type prediction \\
        (2) Next token-level prediction \\
        (3) statement-level prediction \\
        }} & N/A & Code & AST \\
        \hline
    \end{tabular}
    }

\end{table*}

Multi-task learning has been applied in the code completion tasks. 
Table~\ref{tab:mtl_compare} summarizes the difference between PyCoder and the existing multi-task learning code completion.
Similar to the existing work, PyCoder leverages multi-task learning for code completion.
However, there are three aspects that PyCoder is different from prior work, namely, the training/testing mechanism, types of multi-task learning, and the sources of information.

\begin{enumerate}
    \item \textbf{Training/testing mechanism:} Prior studies~\cite{liu2020self, liu2022unified, izadi2022codefill} generally used code+AST for both fine-tuning and inference. As mentioned in the Motivation section, AST information requires the completeness of source code, limiting its practical application in various real-world scenarios. Instead, PyCoder leverages code+token type in the training, but only code for the inference, making our PyCoder can perform on-the-fly code completion, while existing AST-based code completion cannot.
    \item \textbf{Types of multi-task learning:} Similar to prior studies~\cite{liu2020multi, liu2020self, liu2022unified, izadi2022codefill}, our PyCoder leverages multi-task learning for code completion. However, prior studies only investigated a few types of multi-task learning (e.g., hard parameter sharing alone or with soft parameter sharing). Instead, this paper conducted a systematic comparison to confirm which types of multi-task learning perform the best (see RQ2). We found that hard parameter sharing performs best, highlighting the novelty of the finding that existing work never explores before. 
    \item \revise{R3.4}{\textbf{The types of syntactic information:}} There are various forms that can be used to represent syntactic information, e.g., AST and standard token types. Prior studies~\cite{liu2020self, liu2022unified, izadi2022codefill} mainly focus on AST information. As discussed in the Motivation section, such AST token types are formal and required the completeness of source code, limiting their applicability. Instead, we are the first to leverage the standard token type information, which (1) is more static, abstract, and lightweight information, (2) follows the natural order of code sequences, and (3) can be extracted at any time without requiring the completeness and syntactically correct of the code snippets as AST does. These benefits allow PyCoder to perform on-the-fly code completion, while being syntax-aware.
\end{enumerate}

Based on the empirical results from RQ1, we confirm that PyCoder performs the best. 
In addition, the ablation studies from RQ2-4 also confirm that the design architecture of our PyCoder plays a significant role in performance improvement, highlighting the significant advancement and novelty of our contributions to the code completion literature.

}

\section{Threats to Validity}
\label{sec:threats to validity}

% In this section, we discuss threat to the validity.

% relate to the granularity level of code completion, and the baseline selection.
% Our \our~is trained to work on token-level code completion and can be adapted to line-level code completion.
% Thus, the model is generalized to use in different granularity.

\textbf{Threats to construct validity} relate to the selection of baseline approaches.
In this paper, we select the publicly accessible approach, which could reduce biases and increase the transparency of the comparison of the experimental results. 
Therefore, 
% instead of comparing with nonpublic MTL code completion models~\cite{izadi2022codefill, liu2020self, liu2020multi}, 
we select the competitive state-of-the-art approaches which are publicly available by the authors as the baselines.
We run all the experiments using the replication package and the best hyperparameter settings in their papers.

\revise{R3.7}{
Additionally, regarding the computational cost, there will be some additional computation costs related to token type data extraction and the model training time.
However, the inference cost remains the same as the state-of-the-art models like CodeGPT and GPT-2 that PyCoder built upon (i.e, same model size, model parameters, inference time). 
Given the fact that the model is trained once based on a snapshot of datasets without the need for model retraining, the additional cost will not directly impact the end users and should not be a major concern. 
}

\revise{R2.3}{
\textbf{Threats to internal validity} relate to the impact of the hyperparameters on the performance of \our.
To mitigate this threat, we conduct experiments with various hyperparameter settings (see RQ3 and RQ4).
However, we find that \our~is generally robust to the model task weights.
Thus, we suspect that hyperparameters will have a minimal impact on the performance of \our.
Nevertheless, optimizing the hyperparameters of the Transformer model could be expensive and is not the main goal of this paper.
Due to the limited access to premium GPU computing resources, our results serve as a minimum bound, which could be further improved after optimization and with premium GPU access.
Nevertheless, to mitigate this threat, we report the hyperparameter settings in our replication package.}

% These hyperarameters haven't be optimized due to the very expensive cost in large search space of Transformer architecture.
% Thus, there might be room for further improvement; nonetheless, the current settings have the considerable performance to compete with the baselines.
% To mitigate this threat, we reports the hyperarameter settings for future replication studies.

\revise{R2.6}{
\textbf{Threats to external validity} relate to the generalizability of our approach.
The evaluation of our approach is limited to the PY150 dataset, where the testing set consists of 50,000 python files.
The PY150 dataset is a standard benchmark dataset for code completion, which has been used in prior studies and Microsoft's CodeXGlue~\cite{lu2021codexglue, kim2021code, izadi2022codefill, li2017code, liu2020self, liu2022unified, wang2020towards}, ensuring a fair comparison of our work with the prior studies.
However, the results may not be generalized to other programming languages, projects and contexts.
Although we could potentially collect a larger size of the dataset, a comparison of our approach with our own collected dataset may pose various potential threats to validity, e.g., obtaining different results reported in CodeXGLUE, unfair comparison with the existing work, etc.}

\revise{R2.7}{
In addition, our results are limited to the Python programming language only.
Nevertheless, it is also possible to extend our approach to other programming languages.
For example, researchers can apply Java's tokenizer library to extract the token type information. 
For Java, researchers could use the standard “javalang” library\footnote{ https://github.com/c2nes/javalang/blob/master/javalang/tokenizer.py}.
Therefore, other languages can be explored in the future.
}

\revise{R2.9}{
Finally, our evaluation is limited to line-level and token-level accuracy measures.
Such measures only evaluate the model performance (which is the goal of this work), but it does not reflect user satisfaction and its impact on developer productivity. 
This is also the key important limitation that applies to the existing code completion studies. 
To answer the research question (How do the right/wrong predictions of PyCoder impact developer productivity?), we believe that an actual tool must be developed and must be used by actual developers. 
Unfortunately, PyCoder is still at the early stage of development, not yet ready to be a prototyping tool, preventing us to conduct a rigorous and comprehensive analysis of how the predictions from PyCoder impact developer productivity. 
Then, human-centric research must be used, for example, an observational study, an intervention study, and an ethnographic study. 
Given the fact that this is an open-challenge research question that requires specific research methodologies, we suggest that multi-disciplinary research (i.e., human-centered computing, AI, and Software Engineering) is required in order to address this challenging research question.
}

% our experiments are on the synthetic dataset (i.e., PY150) to measure the model performance.
% While in code completion studies, there are still open research challenges to evaluate the performance in the real-world scenario. 
% We believe that a case-control study with actual human experimentation is needed.
% As our PyCoder is still in the early stage of development, we leave these challenging experiments to future research.

\section{Conclusion}\label{sec:conclusion}

In this work, we propose \our~to leverage token types, a kind of lightweight syntactic information, with a multi-task training strategy that learning on the supporting task of predicting token types during the training phase.
We intensively train and test our \our~on different multi-task training techniques, task weighing parameters, and decoding methods to find the best suitable architecture.
Our study underline the following conclusion:
\begin{itemize}
    \item \our~surpasses all the state-of-the-art models in our setting and also receives the first place in CodeXGLUE’s python code completion benchmark. The results indicate that the token type syntactic information can be beneficial in code completion.
    \item In our setting, MTL: Hard Parameter Sharing -- PyCoder-Hard with task's weight (Type:Code) 1:9 and Beam Search performs the best. 
    \item  Our study highlights the importance of investigating various choices of setting (e.g., multi-task training strategies, parameter setting) instead of solely relying on suggestions from prior work.
\end{itemize}
Our \our~has extended the feature of on-the-fly code completion with lightweight syntactic-aware information.
However, we acknowledge that there is still a space to develop the fully syntactically correct code completion model with on-the-fly feature.
We leave this exploration for the future research study.

\section*{Acknowledgment}
Chakkrit Tantithamthavorn was partly supported by the Australian Research Council’s Discovery Early Career Researcher Award (DECRA) funding scheme (DE200100941).

\bibliography{mybibfile}

% \appendices
% \input{sections/appendix}

% \section{}

% \ifCLASSOPTIONcompsoc
%   % The Computer Society usually uses the plural form
%   \section*{Acknowledgments}
% \else
%   % regular IEEE prefers the singular form
% \fi

\ifCLASSOPTIONcaptionsoff
  \newpage
\fi

\bibliographystyle{IEEEtran}

\begin{IEEEbiography}[{\includegraphics[width=1in,height=1.25in,clip,keepaspectratio]{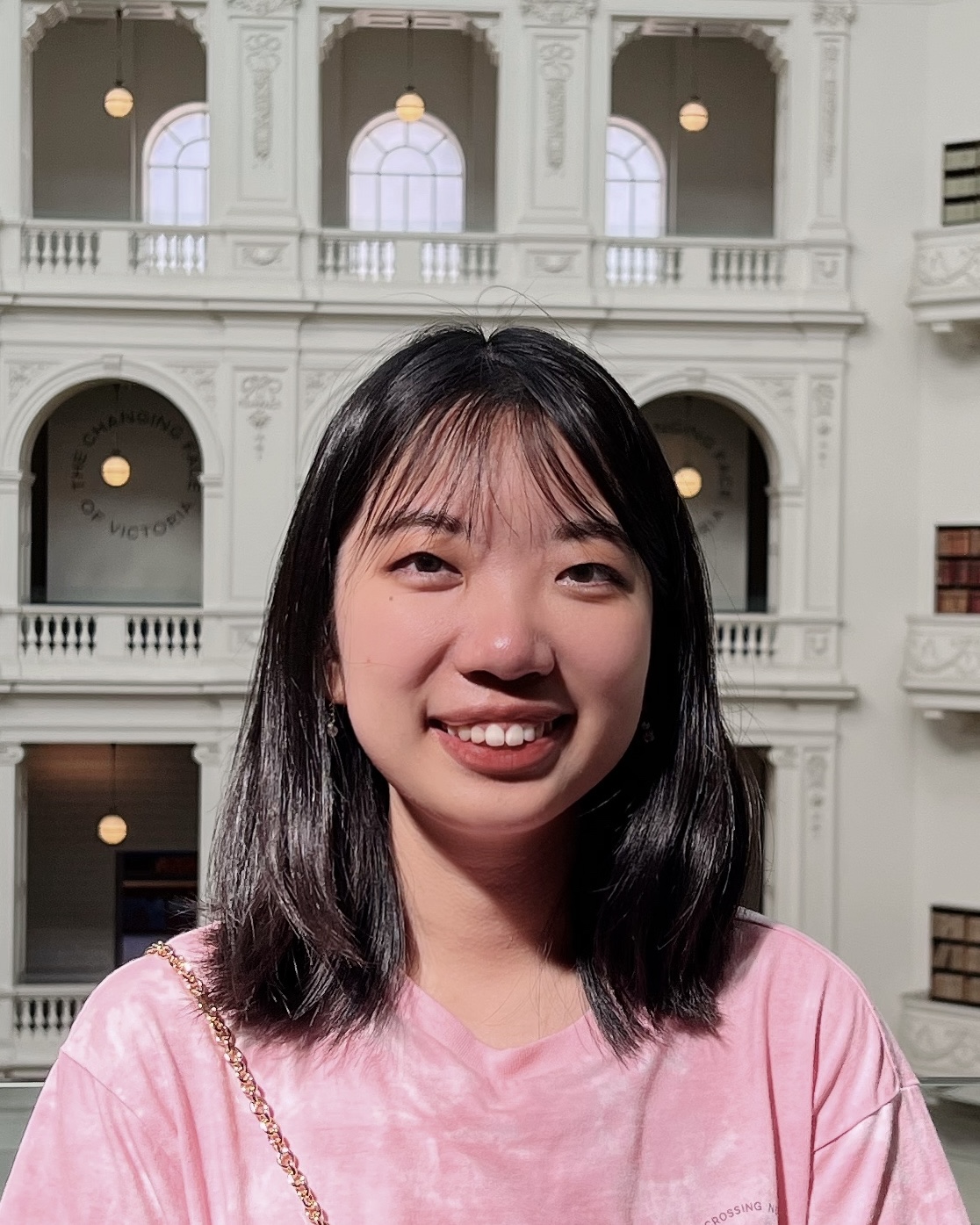}}]{Wannita Takerngsaksiri}
is a Ph.D. candidate at Monash University, Australia. Her research interest includes code generation, machine learning (ML), and natural language processing (NLP).
Specifically, her research goal aims to consider multiple aspects of code completion and to develop computational methods and advanced AI techniques to assist the coding process to be more effective for developers in practice. 
\end{IEEEbiography}

\begin{IEEEbiography}[{\includegraphics[width=1in,height=1.25in,clip,keepaspectratio]{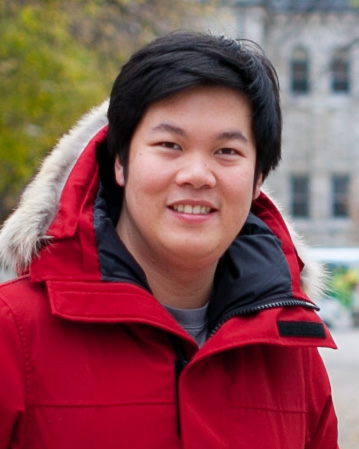}}]{Chakkrit (Kla) Tantithamthavorn}
is an ARC DECRA Fellow and a Senior Lecturer in Software Engineering in the Faculty of Information Technology, Monash University, Australia. He is pioneering an emerging research area of Explainable AI for Software Engineering (http://xai4se.github.io), inventing many AI-based technologies to improve developers’ productivity and make software systems more reliable and more secure, while being explainable to practitioners. To date, the XAI4SE book has attracted 10,000+ page views from 70 countries worldwide. He regularly published at TSE, ICSE, FSE, EMSE, ASE, and MSR, all of which are top software engineering venues. The excellence of his research is recognized through many awards including an ACM SIGSOFT Distinguished Paper Award 2021, an ARC’s Discovery Early Career Researcher Award 2020, the World Most Impactful Early-Stage SE Researcher based on a bibliometric assessment of software engineering (2013-2020). 
\end{IEEEbiography}

\begin{IEEEbiography}[{\includegraphics[width=1in,height=1.25in,clip,keepaspectratio]{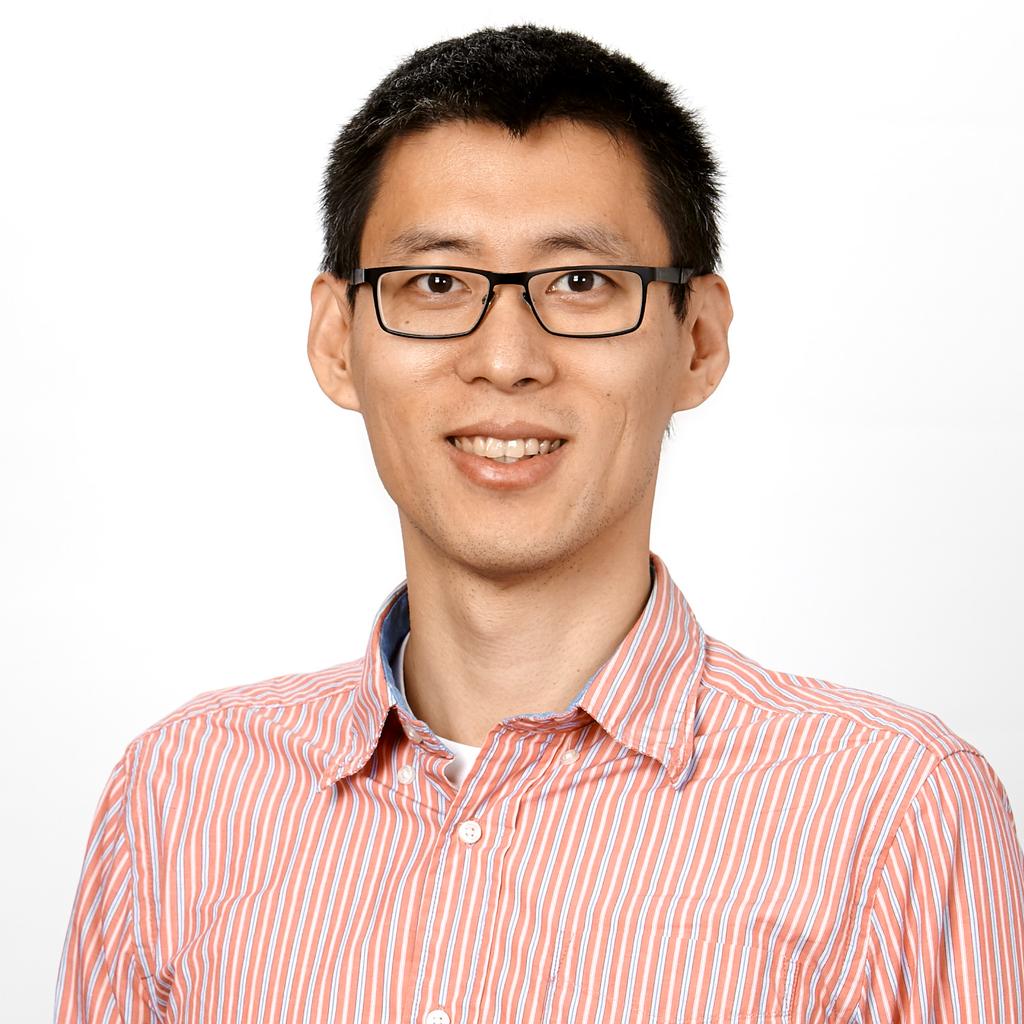}}]{Yuan-Fang Li}
is an Associate Professor at Faculty of IT, Monash University. His research interest is artificial intelligence, particularly the intersection between natural language processing and knowledge representation. His recent investigations include the following tasks: (1) neuro-symbolic approaches to complex question answering, (2) knowledge graph construction from text/images, and (3) graph representation learning. His research work has been published at top AI and NLP venues including ACL, EMNLP, ECCV, ICCV, ICLR, NeurIPS, IEEE TNNLS, and Pattern Recognition. 
\end{IEEEbiography}

\end{document}